\documentclass[runningheads]{llncs}
\usepackage[T1]{fontenc} % T1 fonts for proper output in PDFs

\usepackage{graphicx} % For including graphics
\usepackage{caption}  % For customizing captions
\usepackage{subcaption} % For subfigures
\usepackage{float}    % For precise figure placement
\usepackage{tikz} % For creating vector graphics and diagrams

\usepackage{lipsum} % For generating dummy text
\usepackage{textcomp} % Additional text symbols
\usepackage{xspace} % For managing space after macros

\usepackage{amsmath,amsfonts}
\usepackage{amssymb}% Standard math environments and fonts
\usepackage{bm} % Bold math symbols

\usepackage[ruled,vlined]{algorithm2e}% Algorithm environment and commands
\usepackage[utf8]{inputenc}
\usepackage{multirow} % Multi-row cells in tables
\usepackage{makecell} % Customized table cells

\usepackage{balance} % For balancing columns on the last page
\usepackage[switch]{lineno} % Line numbering (switch option for two-column mode)

\usepackage{hyperref} % For hyperlinks
\usepackage[symbol]{footmisc} % Footnotes with symbol markers

\newcommand{\ouralg}{$\mathsf{TARDIS}$\xspace}
\begin{document}
\title{Power-Aware Scheduling for Multi-Center HPC Electricity Cost Optimization}

\titlerunning{GNN-Driven Power-Aware Scheduling for
Multi-Center HPC Cost Optimization}

\author{Abrar Hossain\inst{1} \and
Abubeker Abdurahman\inst{2} \and
Mohammad A. Islam\inst{3} \and
Kishwar Ahmed\inst{1}}

\authorrunning{A. Hossain et al.}

\institute{The University of Toledo, USA \and
Amazon Web Services, USA \and
The University of Texas at Arlington, USA}

% \institute{The University of Toledo, USA \\
% \email{\{abrar.hossain, kishwar.ahmed\}@utoledo.edu} \and
% Amazon Web Services, USA\\
% \email{abukemoh@amazon.com} \and
% The University of Texas at Arlington, USA \\
% \email{mislam@uta.edu}}

{\maketitle}              % typeset the header of the contribution
\begin{abstract}
This paper introduces \ouralg (\textbf{T}emporal \textbf{A}llocation for \textbf{R}esource \textbf{D}istribution using \textbf{I}ntelligent \textbf{S}cheduling), a novel power-aware job scheduler for High-Performance Computing (HPC) systems that minimizes electricity costs through both temporal and spatial optimization. Our approach addresses the growing concerns of energy consumption in HPC centers, where electricity expenses constitute a substantial portion of operational costs and have a significant financial impact. \ouralg employs a Graph Neural Network (GNN) to accurately predict individual job power consumption, then uses these predictions to strategically schedule jobs across multiple HPC facilities based on time-varying electricity prices. The system integrates both temporal scheduling—shifting power-intensive workloads to off-peak hours—and spatial scheduling—distributing jobs across geographically dispersed centers with different pricing schemes. We evaluate \ouralg using trace-based simulations from real HPC workloads, demonstrating cost reductions of up to 18\% in temporal optimization scenarios and 10-20\% in multi-site environments compared to state-of-the-art scheduling approaches, while maintaining comparable system performance and job throughput. Our comprehensive evaluation shows that \ouralg effectively addresses limitations in existing power-aware scheduling approaches by combining accurate power prediction with holistic spatial-temporal optimization, providing a scalable solution for sustainable and cost-efficient HPC operations.

\end{abstract}

\section{Introduction}
\label{sec:intro}

\subsection*{Why is energy consumption a concern in HPC?}

High-Performance Computing (HPC) systems are essential for scientific research, large-scale simulations, and data-intensive applications, but their growing computational demands have led to significant energy consumption challenges. Modern HPC centers require up to 30 MW of power\cite{inl2024mobile}—comparable to small cities—with some facilities reporting annual electricity costs exceeding  \$40 million~\cite{enterpriseviewpoint2024sustainable}. As the Top500~\cite{top500} most powerful systems consume multiple megawatts continuously~\cite{miyazaki2018bayesian}, and global data center energy usage is projected to reach 3\% of worldwide electricity by 2030~\cite{nana2023energy}, optimizing energy efficiency has become critical for both sustainability and operational costs.
Electricity providers implement dynamic pricing structures that create both challenges and opportunities. On-peak periods incur rates 200-300\% higher than off-peak hours~\cite{borenstein2002dynamic}, with U.S. industrial electricity costs ranging from \$0.06/kWh during off-peak to over \$0.20/kWh during peak demand~\cite{santos2022understanding}. European facilities face similar price volatility exceeding 250\%\cite{yang2013integrating}. Strategically shifting workloads to off-peak hours can reduce power costs by up to 40\%\cite{liu2013data} while aligning with renewable energy availability can decrease carbon emissions by 10-15\%\cite{haghshenas2020infrastructure}. Research demonstrates that demand-response scheduling can achieve 20-40\% cost reductions without significantly impacting performance\cite{yang2013integrating}, and even redistributing just 25\% of workloads to off-peak periods yields substantial savings while maintaining system efficiency~\cite{purkayastha2018holistic}.

\subsection*{What are the challenges in optimizing HPC job scheduling for energy efficiency?}

Traditional job scheduling algorithms (e.g., First Come First Serve (FCFS), Backfilling) and SLURM-based scheduling focus primarily on optimizing resource allocation and job throughput~\cite{fan2021job}. While these approaches are effective for maximizing system utilization and reducing job wait times, they fail to account for the energy consumption variability across different workloads. Modern schedulers have started incorporating energy-awareness strategies, such as Dynamic Voltage and Frequency Scaling (DVFS) and power-aware job migration, but they are often limited by static power capping policies, heuristic-based energy models, or basic workload profiling that do not fully leverage dynamic electricity pricing. Additionally, these methods lack a robust mechanism to predict power consumption at a fine-grained level, leading to inefficient energy utilization and missed opportunities for cost optimization~\cite{fan2021job}. One of the key challenges in power-aware job scheduling is accurately predicting the power consumption of individual jobs~\cite{antici2023pm100}. Factors such as job type, resource usage, workload intensity, and system heterogeneity can significantly impact power consumption~\cite{borghesi2017power}. Studies by Halder et al.~\cite{halder2024empirical} have shown that incorporating machine learning models can improve power prediction accuracy by up to 35\%, with deep learning techniques demonstrating even greater potential. According to Saillant et al.~\cite{saillant2020predicting}, workload-aware power forecasting can improve energy efficiency by up to 30\% compared to static models. They also indicate that real-time power prediction models can lead to 20--50\% better scheduling efficiency, reducing overall energy waste in HPC systems. However, existing schedulers often lack the integration of such sophisticated forecasting techniques, leading to suboptimal energy scheduling and inefficient power management.

\subsection*{What are the limitations of existing approaches?}

Most existing power-aware schedulers operate with simplified power prediction models that fail to capture the complex relationship between job characteristics and power consumption. Zhou et al. \cite{zhou2014reducing} used a static model for Blue Gene/P systems that simply divides power consumption between working and idle nodes without considering workload-specific variations. Similarly, Yang et al. \cite{yang2013integrating} relied on historical data based on user ID and project information, assuming job repeatability without addressing the dynamic nature of scientific workloads. Another significant limitation lies in the optimization scope. Existing approaches predominantly focus on temporal optimization alone, disregarding spatial dimensions. Both Zhou \cite{zhou2014reducing} and Yang \cite{yang2013integrating} implemented window-based scheduling with 0-1 knapsack formulations that only consider on-peak and off-peak periods within a single facility. Patki et al. \cite{patki2015practical} introduced RMAP with fair-share power allocation and overprovisioning but limited to a single cluster environment. These approaches cannot leverage the potential cost savings from geographical electricity price variations. Current methods also suffer from rigid power capping mechanisms that compromise performance. Mammela et al. \cite{mammela2012energy} implemented energy-aware scheduling algorithms that power off idle nodes but lacked dynamic power prediction, while Solorzano et al. \cite{solorzano2024toward} employed fixed "power knobs" on the Fugaku supercomputer that require manual intervention from users. These approaches fail to automatically adapt to workload characteristics and electricity price dynamics. Load balancing approaches such as Pinheiro et al. \cite{pinheiro2001load} focus primarily on turning nodes on/off based on load distribution without considering electricity pricing variations. While they achieve energy savings, they miss opportunities for cost optimization through strategic job placement aligned with price fluctuations. Though Murali et al. \cite{murali2024metascheduling} attempted to address both electricity costs and response times in a compute grid using Minimum Cost Maximum Flow formulation, their approach relied on separate prediction models for response time and electricity prices without an integrated understanding of workload power characteristics. Their SARIMA model for electricity price prediction also cannot capture the complex patterns that emerge from renewable energy integration and market dynamics. Our work addresses these limitations through a unified approach that combines GNN-based power prediction with spatial-temporal optimization across multiple HPC centers. Unlike previous approaches that separate power prediction from scheduling decisions, our framework integrates these components to make holistic optimization decisions that consider both when and where to schedule power-intensive jobs.
% \subsection*{How can Graph Neural Network (GNN) and multi-HPC center scheduling help reduce electricity costs in HPC?}
\subsection*{How we contribute to reducing electricity costs in HPC?}
In this paper we propose TARDIS (Temporal Allocation for Resource Distribution using Intelligent Scheduling), a novel power-aware job scheduler that utilizes a Graph Neural Network (GNN) to predict the power consumption of individual HPC jobs. The scheduler integrates these predictions to simultaneously optimize job placement across both time and geographic location. By analyzing electricity pricing patterns at multiple HPC centers, our approach strategically assigns jobs to minimize overall costs while maintaining performance. This unified spatial-temporal optimization leverages both daily rate fluctuations and regional pricing differences, achieving cost savings impossible with temporal scheduling alone. To evaluate the effectiveness of our proposed scheduler, we employ a trace-based simulation, utilizing historical workload traces from HPC system~\cite{antici2023pm100}. 
% Trace-based simulation offers several advantages, including the ability to replicate realistic system behavior, to assess scheduler performance under diverse workload conditions, and to identify potential bottlenecks without impacting production environments. This method has been effectively utilized in prior studies~\cite{zhou2014reducing,fan2021deep}, where trace-based simulations provided valuable insights into job allocation strategies and their impact on system performance.
Our experimental results demonstrate that the GNN-based scheduler significantly reduces electricity costs compared to state-of-the-art scheduling approaches while maintaining or improving job throughput. Specifically, the scheduler achieves a notable reduction in energy costs, with an 18\% cost reduction in temporal optimization scenarios and 10-20\% savings in multi-site environments. Our approach integrates power-aware scheduling, multi-HPC optimization, and ML-driven power prediction to efficiently manage HPC workloads under dynamic pricing. It lays the foundation for future research on renewable energy integration and real-time adaptive scheduling. The primary contributions of this work include:

\begin{itemize}
    \renewcommand{\labelitemi}{$\bullet$}  % Explicitly set bullet symbol
    \item A novel GNN-based power prediction model that estimates the power consumption of HPC jobs with high accuracy.
    \item A dynamic, price-aware scheduling framework that optimizes job execution based on electricity pricing trends.
    \item A multi-center job scheduling extension that enables cost-efficient workload distribution across multiple HPC facilities.
    \item Comprehensive performance evaluation comparing the proposed scheduler against traditional state-of-the-art approaches.
\end{itemize}

% \subsection*{Research Objectives \& Hypothesis}

% The key objectives of this research are:

% \begin{itemize}
%     \item To reduce the total electricity cost of HPC job execution by leveraging dynamic electricity pricing.
%     \item To improve power consumption prediction accuracy using a GNN model.
%     \item To evaluate the effectiveness of the proposed scheduler compared to existing job scheduling algorithms.
% \end{itemize}

% The underlying hypothesis of this research is that integrating accurate power consumption forecasting with electricity price-aware scheduling will lead to significant cost savings while maintaining or improving job throughput.

% \subsection*{\abrar{Structure of the Paper}}

% The remainder of this paper is structured as follows:

% \begin{itemize}
%     \item Section~2 provides a review of existing research on power-aware job scheduling in HPC environments.
%     \item Section~3 details the proposed methodology, including the design of the GNN model and the scheduling algorithm.
%     \item Section~4 presents the experimental setup and evaluation metrics used to assess the effectiveness of the scheduler.
%     \item Section~5 discusses the results and insights
%     \item Section~6 concludes the paper and outlines future research directions.
% \end{itemize}

\vspace{-2mm}
% \section{Preliminaries}
\label{sec:background}

\section{Background and Challenges}

In this section, we examine energy and job patterns in HPC systems and establish the foundation for our power-aware scheduling approach.

\subsection{Variability in Energy Consumption}

\begin{figure}[H]
    \centering
    \begin{subfigure}[b]{0.45\textwidth}
        \centering
        \includegraphics[width=\textwidth]{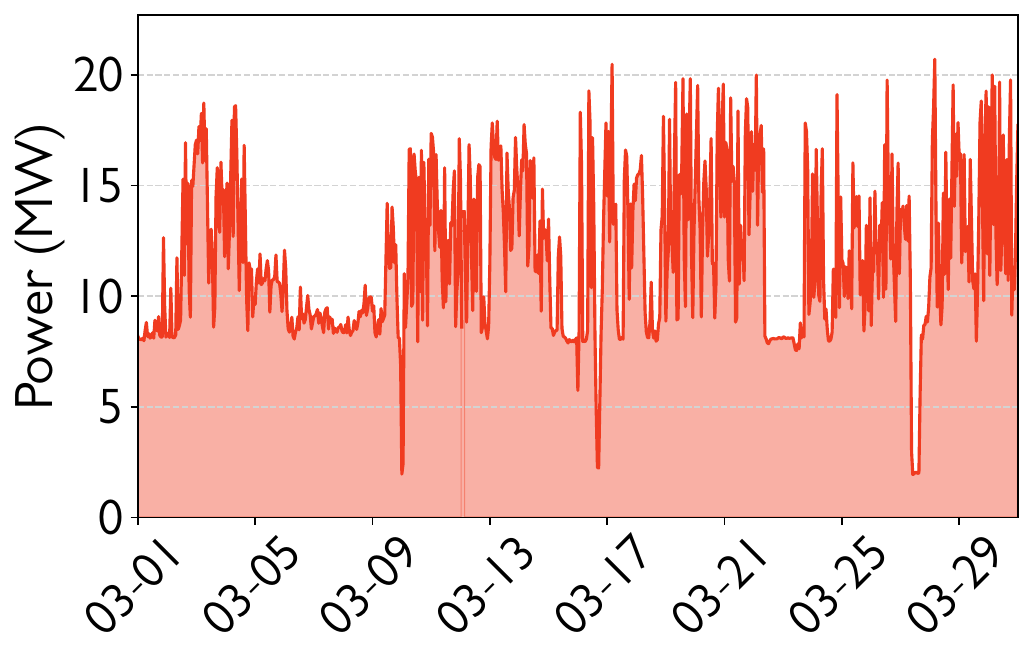}
        \caption{Power consumption.}
        \label{fig:image1a}
    \end{subfigure}
    \hfill
    \begin{subfigure}[b]{0.45\textwidth}
        \centering
        \includegraphics[width=\textwidth]{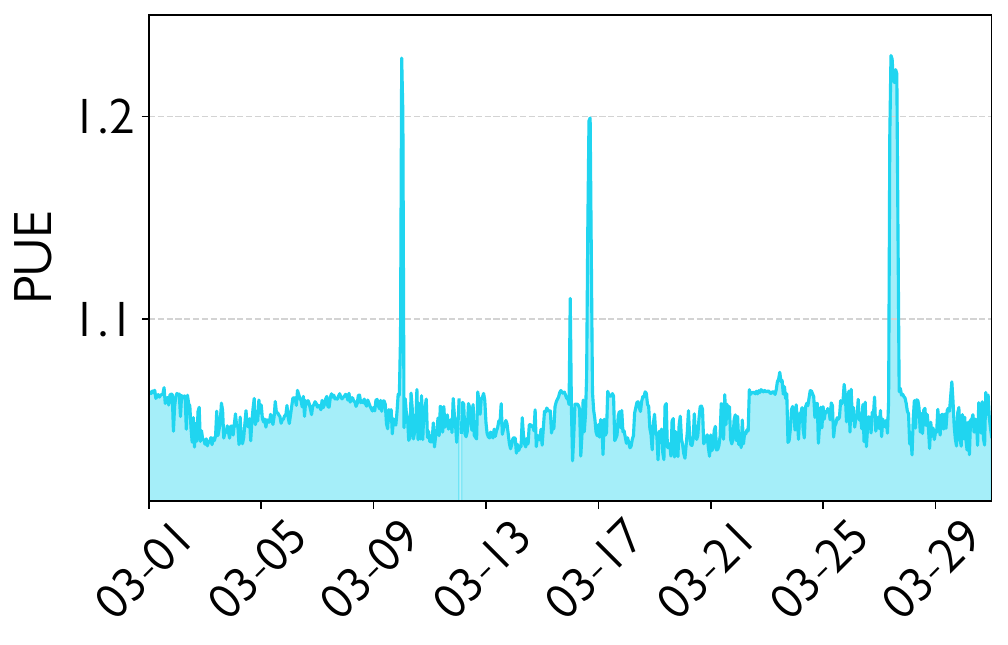}
        \caption{Power usage effectiveness.}
        \label{fig:image1b}
    \end{subfigure}
    \caption{Frontier's power consumption and energy efficiency over March 2023~\cite{sun2024energy}.}
    \label{fig:frontier_analysis}
\end{figure}

HPC systems exhibit significant variations in energy consumption patterns, which directly impact operational costs and efficiency. In Figure~\ref{fig:frontier_analysis}, we present the power consumption data~\cite{sun2024energy} of the Frontier Supercomputer center for the month of March, 2023. The figure highlights the variability in power consumption and Power Usage Effectiveness (PUE) over a typical operational month. Figure~\ref{fig:image1a} demonstrates the power consumption patterns of Frontier's compute nodes. Power usage fluctuates significantly, ranging from idle periods near 2 MW to peaks surpassing 20 MW. This variability is attributed to the dynamic nature of HPC workloads, which depends on job submission rates, computational demands, and resource allocation policies. Such fluctuations highlight inefficiencies in system utilization, where periods of high demand are interspersed with underutilized or idle states. 

Figure~\ref{fig:image1b} illustrates the corresponding PUE. While Frontier achieves a baseline PUE close to 1.1, indicating high energy efficiency, sporadic spikes exceeding 1.2 suggest instances of operational inefficiencies. These spikes may result from system cooling adjustments, non-optimal workload distributions, or transient imbalances in power delivery. Although intermittent, these inefficiencies contribute to significant energy wastage over time.

\subsection{Job Submission Patterns in HPC Systems}

\begin{figure}[H]
    \centering
    \begin{subfigure}[b]{0.45\textwidth}
        \centering
        \includegraphics[width=\textwidth]{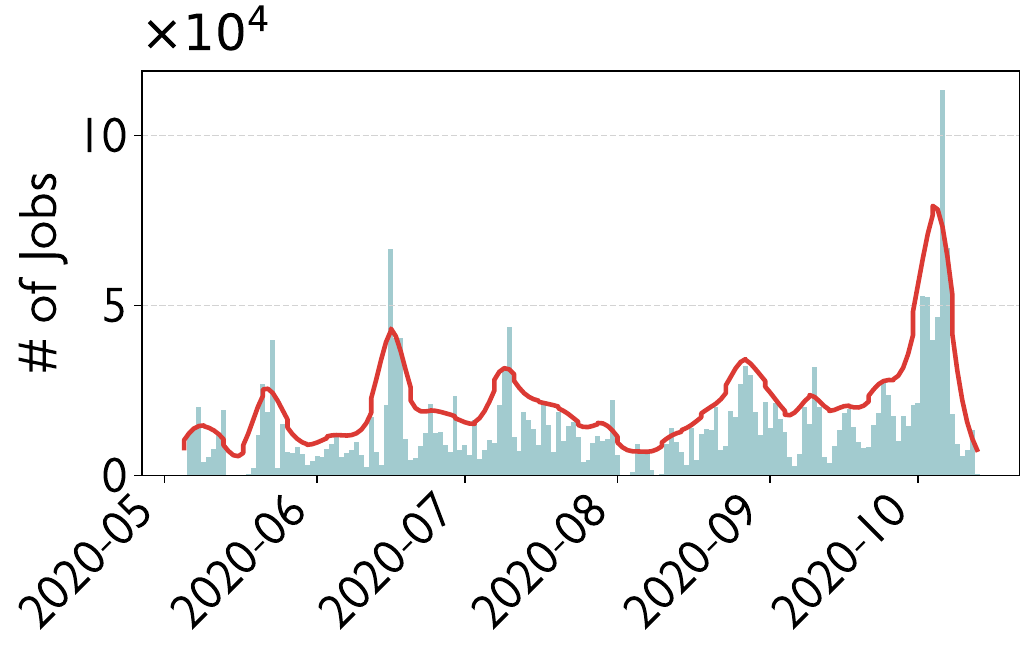}
        \caption{Job submission trends from May - October 2020.}
        \label{fig:job_submission_daily}
    \end{subfigure}
    \hfill
    \begin{subfigure}[b]{0.45\textwidth}
        \centering
        \includegraphics[width=\textwidth]{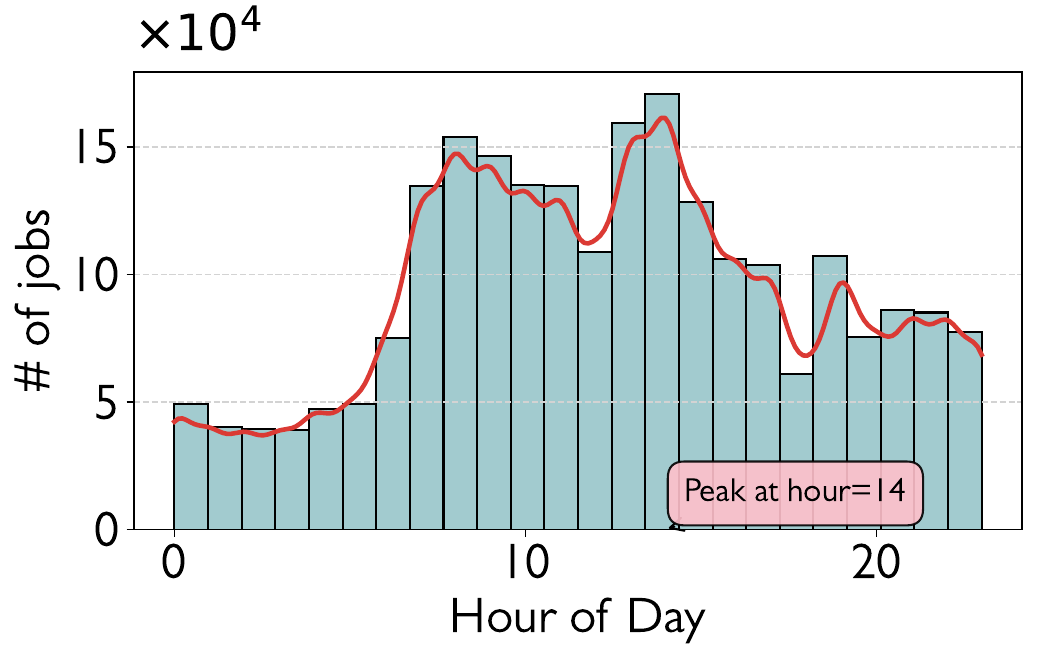}
        \caption{Hourly distribution of job submissions.}
        \label{fig:job_submission_hourly}
    \end{subfigure}
    \caption{HPC job submission patterns over time.}
    \label{fig:job_submission_patterns}
\end{figure}

Energy consumption variability in HPC systems is driven by dynamic job submission patterns. Analysis of Marconi100 job data~\cite{antici2023pm100} (Figures~\ref{fig:job_submission_daily} and~\ref{fig:job_submission_hourly}) highlights this unpredictability. A time-series view and Kernel Density Estimation (KDE) reveal fluctuating daily submissions (May–Oct 2020) with no clear trend or seasonality, except for a peak in October, likely due to ad hoc demands. Hourly job submissions also show irregularity, with a midday peak at 2 PM but no consistent pattern. These findings underscore the need for dynamic scheduling to handle unpredictable workload fluctuations efficiently.

\subsection{Power Consumption Patterns Across HPC Components}

\begin{figure}[H]
    \centering
    \begin{subfigure}[b]{0.45\textwidth}
        \centering
        \includegraphics[width=\textwidth]{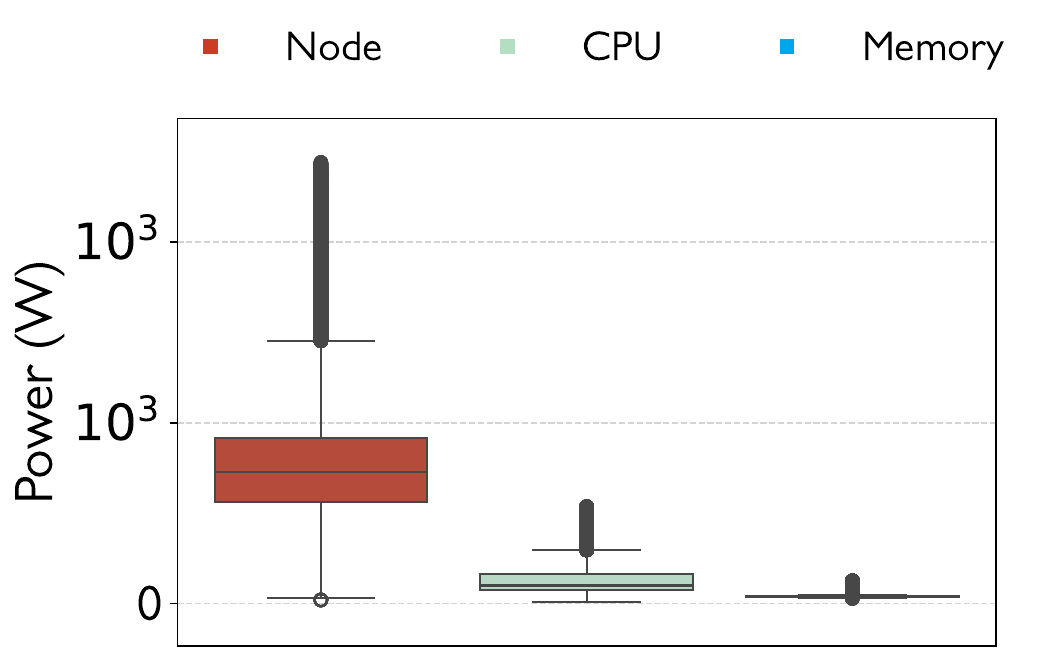}
        % \caption{Power consumption distributions across compute nodes, CPUs, and memory components.}
        \caption{HPC components.}
        \label{fig:component_power}
    \end{subfigure}
    \hfill
    \begin{subfigure}[b]{0.45\textwidth}
        \centering
        \includegraphics[width=\textwidth]{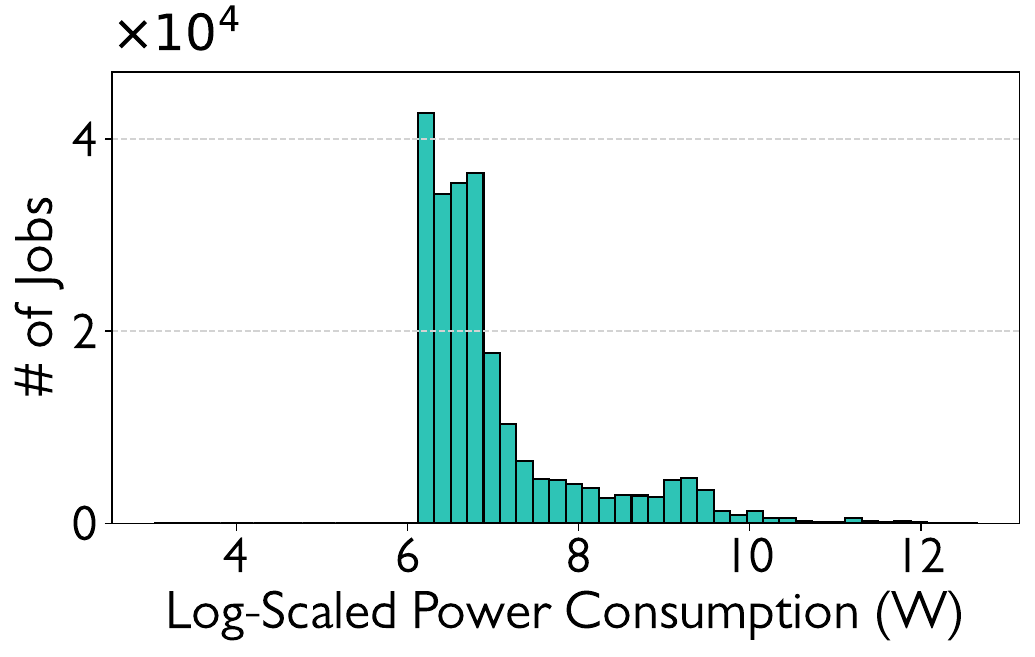}
        % \caption{Power consumption on a logarithmic scale. Long tail toward higher consumption values.}
        \caption{HPC jobs.}
        \label{fig:job_power_distribution}
    \end{subfigure}
    \caption{Power consumption patterns across HPC components and HPC jobs.}
    \label{fig:component_vs_job_power}
\end{figure}

Understanding HPC power consumption is key to optimizing energy use and efficiency. Figure~\ref{fig:component_power} shows a box plot of power distribution across nodes, CPUs, and memory. Compute nodes have the highest variability and peak power, exceeding 2000 W, due to high-power components like GPUs. CPUs show stable power usage with lower variability, while memory consumes the least power with minimal deviation. Since nodes dominate power consumption, they are the primary target for optimization. Node-level power metrics capture aggregate effects from all subcomponents, aiding power-aware scheduling. Studies confirm a strong correlation between CPU usage and power consumption, validating its use as a power proxy~\cite{barroso2007case,fan2007power}.

Figure~\ref{fig:job_power_distribution} illustrates job-level power consumption using a histogram with a logarithmic scale for power values. The distribution reveals that most jobs consume power within the lower ranges, concentrated between \(10^4\) and \(10^6\) W. However, a long tail of jobs extends toward higher power usage, indicating the presence of resource-intensive workloads. These high-power jobs, while fewer in number, contribute significantly to the overall energy consumption of the system. This distribution emphasizes the importance of dynamically managing power-hungry jobs to mitigate their impact on system-wide energy efficiency.

\subsection{Job Exit State Trends}
\begin{figure}[H]
    \centering
    \begin{subfigure}[b]{0.45\textwidth}
        \centering
        \includegraphics[width=\textwidth]{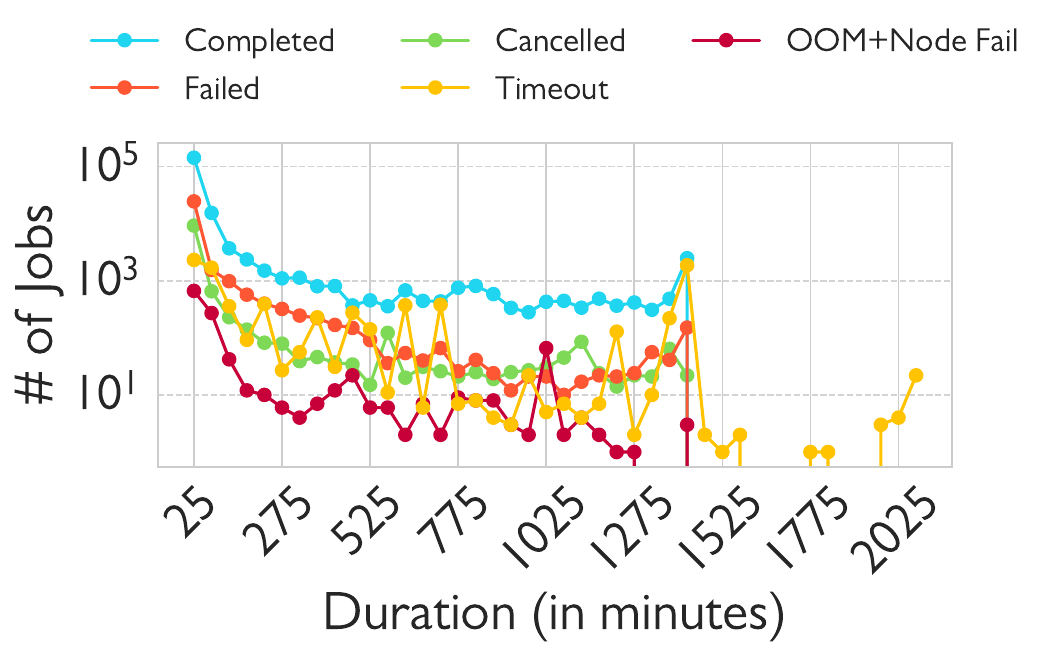}
        \caption{Job trends.}
        \label{fig:exit_state_duration}
    \end{subfigure}
    \hfill
    \begin{subfigure}[b]{0.45\textwidth}
        \centering
        \includegraphics[width=\textwidth]{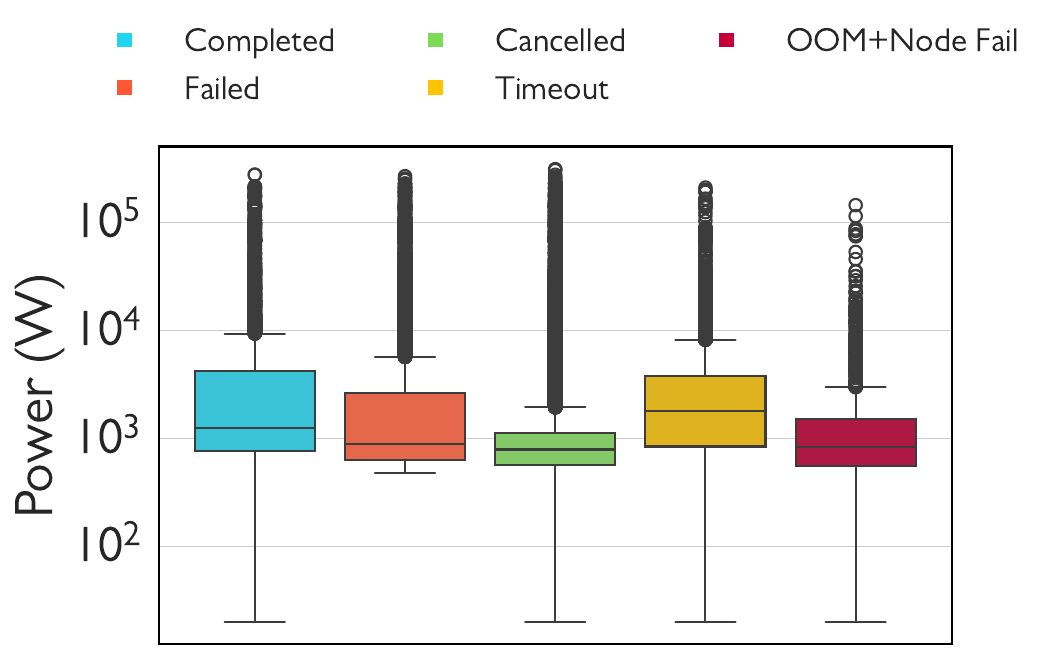}
        \caption{Power consumption.}
        \label{fig:exit_state_power}
    \end{subfigure}
    \caption{Analysis of job trends and power consumption across job exit states.}
    \label{fig:job_exit_state_trends}
\end{figure}

Figure~\ref{fig:exit_state_duration} illustrates how the duration of jobs varies across different exit states, highlighting the distinct power consumption profiles for each state. For example, completed jobs tend to have a more consistent duration, while failed and timeout jobs often exhibit wider variability. This variability suggests inefficiencies in how resources are allocated and jobs are managed within the system. Figure~\ref{fig:exit_state_power} examines the power consumption associated with each job exit state. Completed jobs, while numerous, consume power more efficiently. Failed and timeout jobs, on the other hand, are often associated with higher energy wastage. The figure emphasizes the critical need for energy-aware scheduling that can proactively identify and mitigate inefficiencies linked to specific job states.

\subsection{The Need for Dynamic Scheduling}
The data presented above underscores the pressing need for dynamic job scheduling strategies tailored to reduce energy consumption while maintaining HPC performance. By leveraging intelligent scheduling algorithms, it is possible to smooth power usage fluctuations, optimize thermal management, and minimize the occurrence of high PUE spikes. Dynamic scheduling can: (1) balance system load across unpredictable peaks and troughs, (2) allocate resources more effectively to meet varying job demands, and (3) reduce energy consumption by aligning resource provisioning with actual workload requirements. This study is motivated by the dual challenge of maximizing computational performance and minimizing energy consumption in modern HPC systems. By addressing these challenges, we aim to contribute to sustainable HPC operation practices that align with global efforts to reduce carbon footprints and operational costs.

% \begin{itemize}
%     \item Balance system load across unpredictable peaks and troughs.
%     \item Allocate resources more effectively to meet varying job demands.
%     \item Reduce energy consumption by aligning resource provisioning with actual workload requirements.
% \end{itemize}

% \vspace{-3ex}
\section{Problem Formulation}
\label{sec:prob_form}

We consider a set of batch jobs arriving to a distributed HPC system consisting of $K$ centers over a discrete scheduling horizon of length $T$. Each job $j$ is characterized by a requested node count $s_j$, an estimated runtime $r_j$, and a power usage profile $p_j(t)$ that encodes how much power job $j$ will draw at each time step $t\in\{1,2,\dots,T\}$. Each HPC center $k \in K$ has $N_k$ available nodes and enforces a center-specific power budget $P_{\text{budget},k}$. Electricity prices at each center fluctuate according to $c_k(t)$, which may be positive, zero, or negative in the presence of surplus renewable energy, reflecting local time zones and pricing policies.

The objective is to schedule jobs across both time and space (i.e., different HPC centers) to minimize the total energy cost over the horizon, subject to power and node capacity constraints at each center. We define a binary decision variable $x_{j,k,t}$ that equals 1 if job $j$ is actively running at center $k$ at time $t$, and 0 otherwise. The instantaneous power consumption at HPC center $k$ at time $t$ is given by:
\[
\sum_{j} \Bigl(x_{j,k,t} \cdot p_j(t)\Bigr).
\]

This aggregate power consumption cannot exceed the center-specific budget $P_{\text{budget},k}$, imposing the constraint:
\[
\sum_{j} \Bigl(x_{j,k,t} \cdot p_j(t)\Bigr) \leq P_{\text{budget},k}, 
\quad \forall\,k\in K,\; \forall\,t\in\{1,\dots,T\}.
\]

Similarly, the node capacity constraint guarantees that at any time $t$, the total number of nodes allocated to running jobs remains within each HPC center's limit:
\[
\sum_{j} \Bigl(x_{j,k,t} \cdot s_j\Bigr) \leq N_k, 
\quad \forall\,k\in K,\; \forall\,t\in\{1,\dots,T\}.
\]

To ensure jobs are not split across HPC centers, we enforce that each job must be assigned to at most one HPC center at any time:
\[
\sum_{k\in K} x_{j,k,t} \leq 1, 
\quad \forall\,j,\; \forall\,t\in\{1,\dots,T\}.
\]

Because electricity prices vary with both time and location, the cost incurred at time $t$ at HPC center $k$ is governed by the rate $c_k(t)$. Over one scheduling interval $\Delta t$, the cost contribution of each active job is multiplied by its power usage and the prevailing price at its assigned HPC center. The total energy cost over the entire horizon across all HPC centers is approximated as:
\[
\sum_{k\in K}\sum_{t=1}^{T}
\Bigl(
 c_k(t) \times 
\sum_{j}\bigl[x_{j,k,t}\cdot p_j(t)\bigr]
\Delta t
\Bigr).
\]

The complete optimization problem can be formulated as:

\begin{align*}
\text{Minimize:} \quad & \sum_{k\in K}\sum_{t=1}^{T}
\Bigl(
 c_k(t)\times
\sum_{j}\bigl[x_{j,k,t}\cdot p_j(t)\bigr]
\Delta t
\Bigr), \\
\text{subject to:} \quad & \sum_{j} \bigl[x_{j,k,t} \cdot s_j\bigr] \leq N_k, 
\quad \forall\,k,t, \\
& \sum_{j} \bigl[x_{j,k,t} \cdot p_j(t)\bigr] \leq P_{\text{budget},k}, 
\quad \forall\,k,t, \\
& \sum_{k\in K} x_{j,k,t} \leq 1, 
\quad \forall\,j,t, \\
& x_{j,k,t} \in \{0,1\}, \quad \forall\,j,k,t.
\end{align*}

The above formulation seeks a time- and location-indexed allocation of HPC jobs (through the decision variables $x_{j,k,t}$) that drives high-power phases of computation toward intervals and locations when $c_k(t)$ is minimized or negative, thus reducing overall energy expenditures. 
% At the same time, it must ensure that neither the node capacity nor the instantaneous power budget is violated at any center.

% \subsection*{Scheduling Strategy}

% To preserve system utilization and fairness while exploiting periods of lower (or negative) electricity prices, we propose a window-based scheduling mechanism. Instead of strictly adhering to FCFS order, a scheduling window is defined at each decision time, comprising a group of jobs near the front of the wait queue. Jobs within this window are given equal priority for allocation.

% The scheduler seeks to maximize node usage during inexpensive energy intervals. For example, high-power jobs may be deferred until a negative price window begins, provided such deferral does not violate overarching system policies or lead to extreme unfairness. Thus, the core problem is to determine the assignment of jobs \( \{x_{j,t}\} \) across time \( t \), minimizing the total energy cost while satisfying constraints on node capacity, power budget, and fairness.

\section{TARDIS - Temporal Allocation for Resource Distribution using Intelligent Scheduling}
\label{sec:alg}

\begin{figure}[H]
    \centering
    \includegraphics[width=0.9\textwidth]{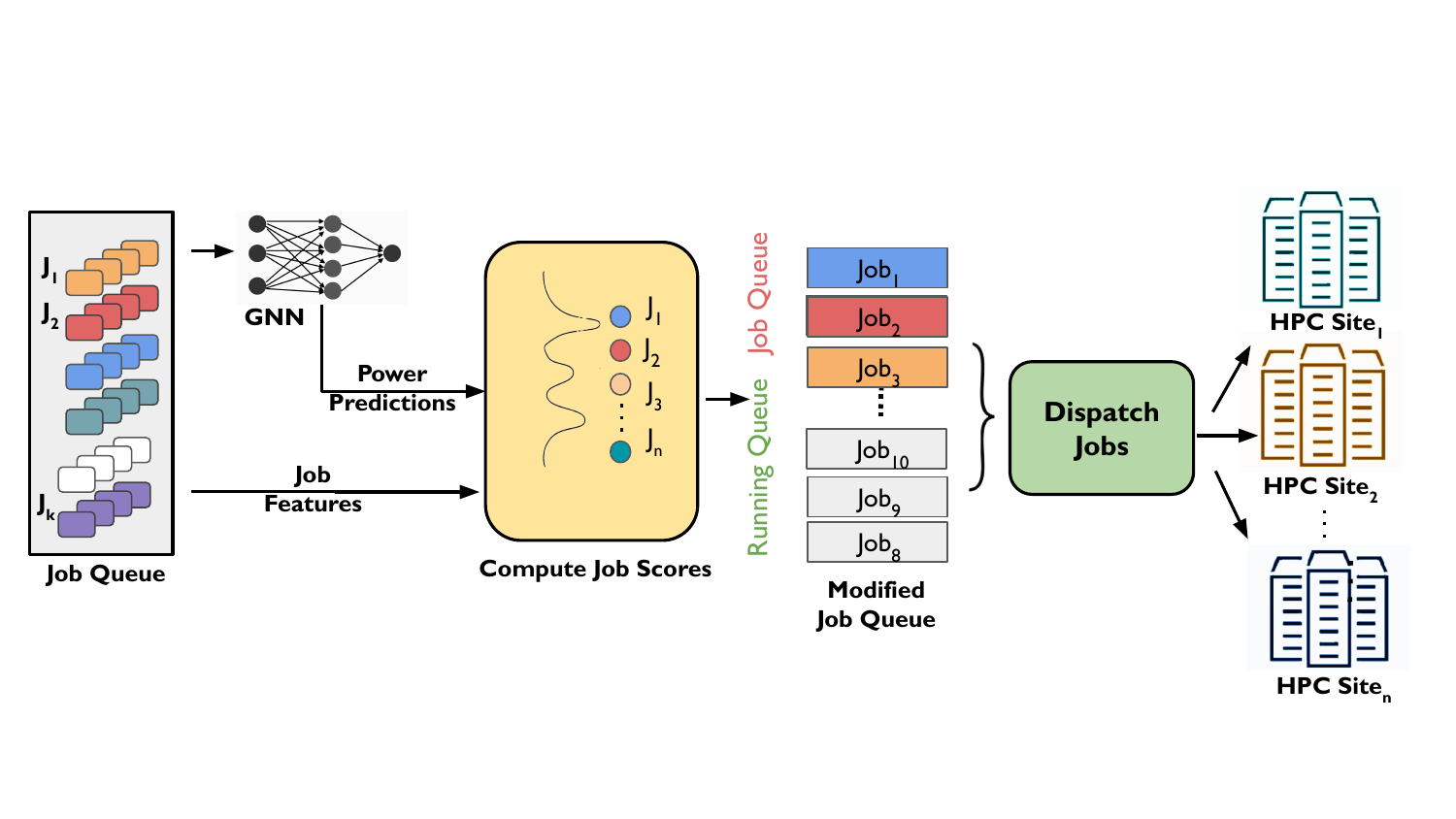}
    \caption{Block diagram of \ouralg.}
    \label{fig:block_diagram}
\end{figure}

The proposed framework \ouralg addresses the challenge of minimizing electricity costs in large-scale HPC systems by combining Graph Neural Network (GNN) based power prediction with a multi-objective scheduling approach that optimizes both temporal and spatial placement of jobs across multiple HPC sites. The overall workflow is illustrated in Figure~\ref{fig:block_diagram}. It consists of three main components: (1) power consumption prediction using GNN, (2) job scoring and queue modification, and (3) spatio-temporal job dispatch.
% \vspace{-2cm}

\subsection{Power Consumption Prediction using GNN}
\ouralg addresses the challenge of minimizing electricity costs in large-scale HPC systems with time-varying energy prices by employing a GNN to predict per-job power consumption. Let \(\{j_1, j_2, \dots, j_N\}\) be the set of submitted jobs over a given time horizon, where each job \(j\) is characterized by a feature vector \(\mathbf{x}_j \in \mathbb{R}^d\). These features include node count, cores per task, cores per node, shared resource flags, priority levels, memory requirements, estimated runtime, and job type encodings. To capture the relational structure among jobs, a k-nearest neighbor approach constructs an undirected graph \(\mathcal{G} = (\mathcal{V}, \mathcal{E})\), where each vertex \(v_j \in \mathcal{V}\) corresponds to job \(j\), and an edge \((v_j, v_{j'}) \in \mathcal{E}\) is introduced if job \(j'\) is one of the \(k\)-nearest neighbors of job \(j\) in the feature space. The feature space is standardized through label encoding of categorical features and normalization of numerical features. After standardizing the feature vectors, the resulting graph-structured mini-batches serve as input to the GNN.

\begin{table}[h]
    \centering
    \renewcommand{\arraystretch}{1.2}
    \begin{tabular}{|l|c|}
        \hline
        \textbf{Component} & \textbf{Power Prediction GNN} \\
        \hline
        \textbf{Input} & $[batch\_size, 8]$ \\
        \hline
        \textbf{Embedding Layer} & Linear$(8 \to 128)$ + BatchNorm + ReLU + Dropout \\
        \hline
        \textbf{GCN Layer 1} & GCNConv$(128 \to 128)$ + BatchNorm + ReLU \\
        \hline
        \textbf{GCN Layer 2} & GCNConv$(128 \to 128)$ + BatchNorm + ReLU + Residual \\
        \hline
        \textbf{FC Layer 1} & Linear$(128 \to 64)$ + ReLU + Dropout \\
        \hline
        \textbf{Output Layer} & Linear$(64 \to 1)$ \\
        \hline
        \textbf{Trainable Parameters} & \textbf{43,265} \\
        \hline
    \end{tabular}
    \caption{Power prediction GNN architecture.}
    \label{tab:gnn_architecture}
\end{table}

\begin{figure}[H]
    \centering
    \begin{subfigure}[b]{0.45\textwidth}
        \centering
        \includegraphics[width=\textwidth]{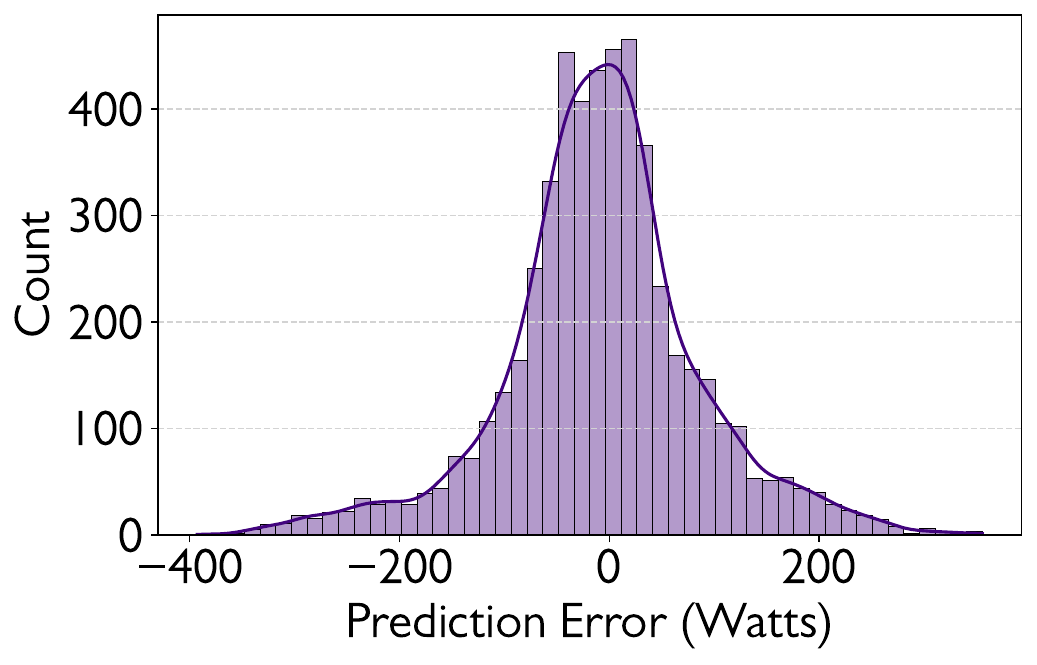}
        \caption{Overall Error distribution.}
        \label{fig:gnn_prediction_error}
    \end{subfigure}
    \hfill
    \begin{subfigure}[b]{0.45\textwidth}
        \centering
        \includegraphics[width=\textwidth]{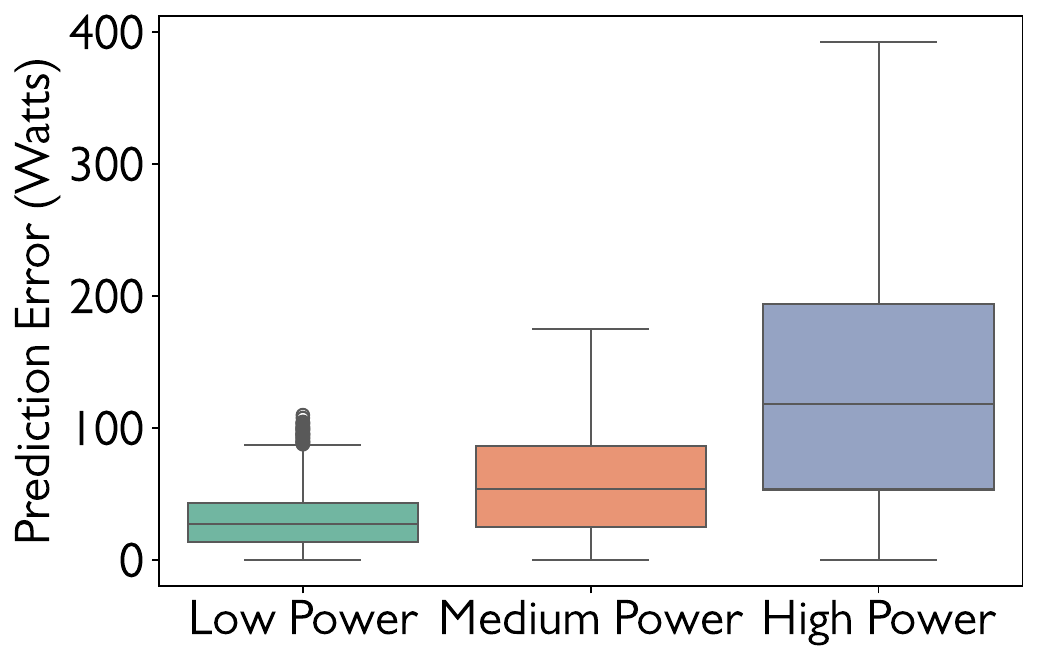}
        \caption{Across three job categories.}
        \label{fig:gnn_job_based_error}
    \end{subfigure}
    \caption{GNN power prediction error distribution.}
    \label{fig:gnn_error}
\end{figure}

The GNN predicts each job's power consumption \(\widehat{P}_j\) through multiple message-passing layers. Let \(\mathbf{h}_j^{(0)} = \mathbf{x}_j\) denote the initial node features. In layer \(\ell\), the hidden state \(\mathbf{h}_j^{(\ell)}\) is updated based on messages from adjacent nodes \(\mathbf{h}_{j'}^{(\ell-1)}\), so that
\begin{equation}
   \mathbf{h}_j^{(\ell)} \;=\; \sigma\!\Bigl(\! W^{(\ell)} \!\sum_{(j', j)\,\in\,\mathcal{E}} \!\alpha_{j',j} \,\mathbf{h}_{j'}^{(\ell-1)}\Bigr),
\end{equation}
where \(W^{(\ell)}\) is a trainable weight matrix, \(\sigma(\cdot)\) is a non-linear activation, and \(\alpha_{j',j}\) is a normalizing factor. The architecture, detailed in Table~\ref{tab:gnn_architecture}, transforms the 8-dimensional input through multiple graph convolution layers with residual connections and batch normalization, ultimately producing a scalar output \(\widehat{P}_j\) representing the predicted power consumption in KW. The GNN model demonstrates robust predictive performance, with prediction errors following an approximately normal distribution centered near zero with a standard deviation of about 100 watts (Figure~\ref{fig:gnn_prediction_error}). While prediction accuracy remains high across job sizes, the model shows increasing error variance for higher-power jobs, with median errors of 30W, 45W, and 120W for low, medium, and high-power jobs respectively (Figure~\ref{fig:gnn_job_based_error}).

\subsection{Multi-Objective Job Scoring}
After obtaining power predictions from the GNN model, we develop a comprehensive scoring mechanism that determines job scheduling priority based on multiple factors. For each job j at site k and time t, the score is computed as:
\begin{equation}
   Score_{j,k,t} = w_c C_{j,k,t} + w_p P_{j,k} + w_u U_{k,t} + w_w W_{j,t}
\end{equation}
The cost factor \(C_{j,k,t}\) incorporates predicted power consumption and time-varying electricity rates:
\begin{equation}
   C_{j,k,t} = \frac{1}{1 + \widehat{P}_j \cdot d_j \cdot c_k(t)}
\end{equation}
where \(\widehat{P}_j\) is the GNN-predicted power consumption, \(d_j\) is the estimated duration, and \(c_k(t)\) represents the electricity rate at site k at time t. The power efficiency factor \(P_{j,k}\) measures the computational efficiency per unit power:
\begin{equation}
   P_{j,k} = \frac{1}{1 + \widehat{P}_j/(N_j \cdot C_j)}
\end{equation}
where \(N_j\) is the number of nodes requested and \(C_j\) is the number of cores per node. The utilization factor \(U_{k,t}\) captures the current resource usage at site k:
\begin{equation}
   U_{k,t} = 1 - \frac{\sum_{j' \in J_k(t)} N_{j'}}{N_k}
\end{equation}
where \(J_k(t)\) represents the set of currently running jobs at site k, and \(N_k\) is the total number of nodes at site k. The wait time factor \(W_{j,t}\) prevents starvation by increasing priority with queue time:
\begin{equation}
   W_{j,t} = \min\left(\frac{t - t_j^{submit}}{T_{max}}, 1\right)
\end{equation}
where \(t_j^{submit}\) is the job submission time and \(T_{max}\) is a normalization constant (typically 24 hours). This formulation ensures that jobs waiting in the queue gradually gain higher priority as they approach the maximum wait threshold, preventing indefinite starvation while still allowing limited delays for power cost optimization. The weighting coefficients \(w_c\), \(w_p\), \(w_u\), and \(w_w\) are empirically tuned to balance the trade-off among energy cost, system efficiency, and job latency.

\subsection{Job Scoring and Queue Modification}
Based on the GNN's power predictions, our scheduling approach uses a multi-criteria scoring mechanism to order and modify the job queue. For each job j at site k and time t, we calculate a priority score that combines cost optimization with traditional scheduling metrics:
\begin{equation}
    Score_{j,k,t} = w_c \cdot C_{j,k,t} + w_p \cdot P_{j,k} + w_u \cdot U_{k,t} + w_w \cdot W_{j,t} + w_r \cdot R_j
\end{equation}
where:
\begin{align*}
    C_{j,k,t} &= \frac{1}{1 + \widehat{P}_j \cdot d_j \cdot c_k(t)} \text{ (cost factor)} \\
    P_{j,k} &= \frac{1}{1 + \widehat{P}_j/(N_j \cdot C_j)} \text{ (power efficiency)} \\
    U_{k,t} &= 1 - \frac{\sum_{j' \in J_k(t)} N_{j'}}{N_k} \text{ (utilization factor)} \\
    W_{j,t} &= \min\left(\frac{t - t_j^{submit}}{T_{max}}, 1\right) \text{ (wait time factor)} \\
    R_j &= \frac{p_j}{p_{max}} \text{ (priority ratio)}
\end{align*}

Here, \(\widehat{P}_j\) is the GNN-predicted power consumption, \(d_j\) is the job duration, \(c_k(t)\) is the electricity rate at site k at time t, \(N_j\) is the requested node count, \(C_j\) is cores per node, \(t_j^{submit}\) is job submission time, and \(p_j\) is the job priority. The weights \(w_c\), \(w_p\), \(w_u\), \(w_w\), and \(w_r\) are empirically tuned to balance energy cost minimization with system performance and fairness.

\subsection{Spatial-Temporal Job Dispatch}
Using the scoring function defined above, the dispatcher optimizes job placement across both time and space dimensions by solving:
\begin{align*}
\text{maximize} \quad & \sum_{j,k,t} Score_{j,k,t} \cdot x_{j,k,t} \\
\text{subject to:} \quad & \sum_{j} x_{j,k,t} \cdot N_j \leq N_k, \quad \forall k,t \\
& \sum_{j} x_{j,k,t} \cdot \widehat{P}_j \leq P_{\text{budget},k}, \quad \forall k,t \\
& \sum_{k} x_{j,k,t} \leq 1, \quad \forall j,t \\
& x_{j,k,t} \in \{0,1\}
\end{align*}

where \(x_{j,k,t}\) is a binary decision variable indicating whether job \(j\) is scheduled on site \(k\) at time \(t\). The constraints ensure that: (1) node capacity limits at each site are respected, (2) power budget constraints at each site are not exceeded, and (3) each job is assigned to at most one site at any given time. This formulation enables the scheduler to simultaneously optimize across both temporal and spatial dimensions, dynamically shifting high-power jobs to times and locations with lower electricity rates while maintaining system constraints.

\section{Evaluation}
\label{sec:evaluation}
In this section we evaluate the performance of \ouralg based on trace based simulation.

\subsection{Experiment Configuration}

\textbf{Job trace:}
We used the PM100 job dataset~\cite{antici2023pm100} for the evaluation of \ouralg. This dataset comprises approximately 230,000 jobs with their corresponding power consumption values. To ensure unbiased evaluation, we used only 30\% of the dataset (approximately 70,000 jobs) for training the GNN model, maintaining temporal ordering in the split. Of these jobs, 80\% were used for training and 20\% for validation. 
% with stratified sampling based on job size and runtime quintiles to ensure representative coverage across different job categories. 
The GNN model was trained using mini-batch gradient descent with the Adam optimizer (learning rate = 0.001) and early stopping with a patience of 15 epochs. We utilized k=5 nearest neighbors for graph construction and incorporated batch normalization and residual connections to enhance model stability. 
% As demonstrated in Figure 5, the model achieves strong predictive performance with errors normally distributed around zero. 
For evaluating the complete scheduling algorithm, we constructed three distinct workload scenarios from the remaining 70\% of the dataset, sampling from different months to test various load conditions. The high workload scenario uses data from July-August, characterized by increased submission rates and higher system utilization. The low workload scenario focuses on October, featuring reduced system load and more intermittent job submissions. The average workload scenario combines data from May-June and September, representing typical operational patterns.

\textbf{Dynamic electricity pricing:}
Our evaluation implements a dynamic pricing model that reflects real-world electricity rate variations. For the \textbf{Site A}, we set a base rate of \$0.12/kWh during off-peak hours (22:00-06:00 EST) and \$0.36/kWh during peak hours (06:00-22:00 EST). The \textbf{Site B} operates with a base rate of \$0.10/kWh during off-peak hours (21:00-05:00 EST) and \$0.30/kWh during peak hours (05:00-21:00 EST). The \textbf{Site C} uses a base rate of \$0.08/kWh during off-peak hours (19:00-03:00 EST) and \$0.24/kWh during peak hours (03:00-19:00 EST). These rate structures are based on typical commercial electricity pricing patterns and time-of-use tariffs observed in major US power distributor locations (East, West, South)\cite{eia2024electricity} . The substantial peak/off-peak differential (3×) was chosen to reflect emerging trends in dynamic electricity pricing and to stress-test the algorithm's ability to optimize across temporal and spatial dimensions.

\textbf{Power budget:}
We implement a power budget constraint that serves as a key mechanism for temporal optimization in our scheduling algorithm. This budget establishes an upper limit on the instantaneous power consumption during peak pricing periods, creating a structured approach to load shifting. The power budget functions in three critical ways: (1) as a hard constraint that prevents the total power draw from exceeding infrastructure limits during peak hours, (2) as a triggering mechanism that identifies and defers high-power jobs to off-peak periods, and (3) as a scoring modifier that dynamically adjusts job priorities based on their power intensity and the system's current load. Rather than merely capping power consumption, our approach uses the power budget as an optimization tool that strategically shifts the execution of power-intensive workloads to periods when electricity rates are lower. This creates an effective balance between immediate job execution and cost optimization, without significantly impacting overall throughput. We configure the power budget at varying percentages (25\%, 50\%, 75\%, and 100\%) of the system's peak historical power consumption to evaluate the algorithm's effectiveness across different operational constraints.

\subsection{Experiment Results}

In this section, we present experiment results from trace-based simulation.

\textbf{Temporal optimization performance:}
We evaluate the temporal optimization performance of \ouralg against three widely-used HPC scheduling policies: First Come First Serve (FCFS), Smallest Job First (SJF), and Backfilling. 

\begin{figure}[H]
    \centering
    \includegraphics[width=1\textwidth]{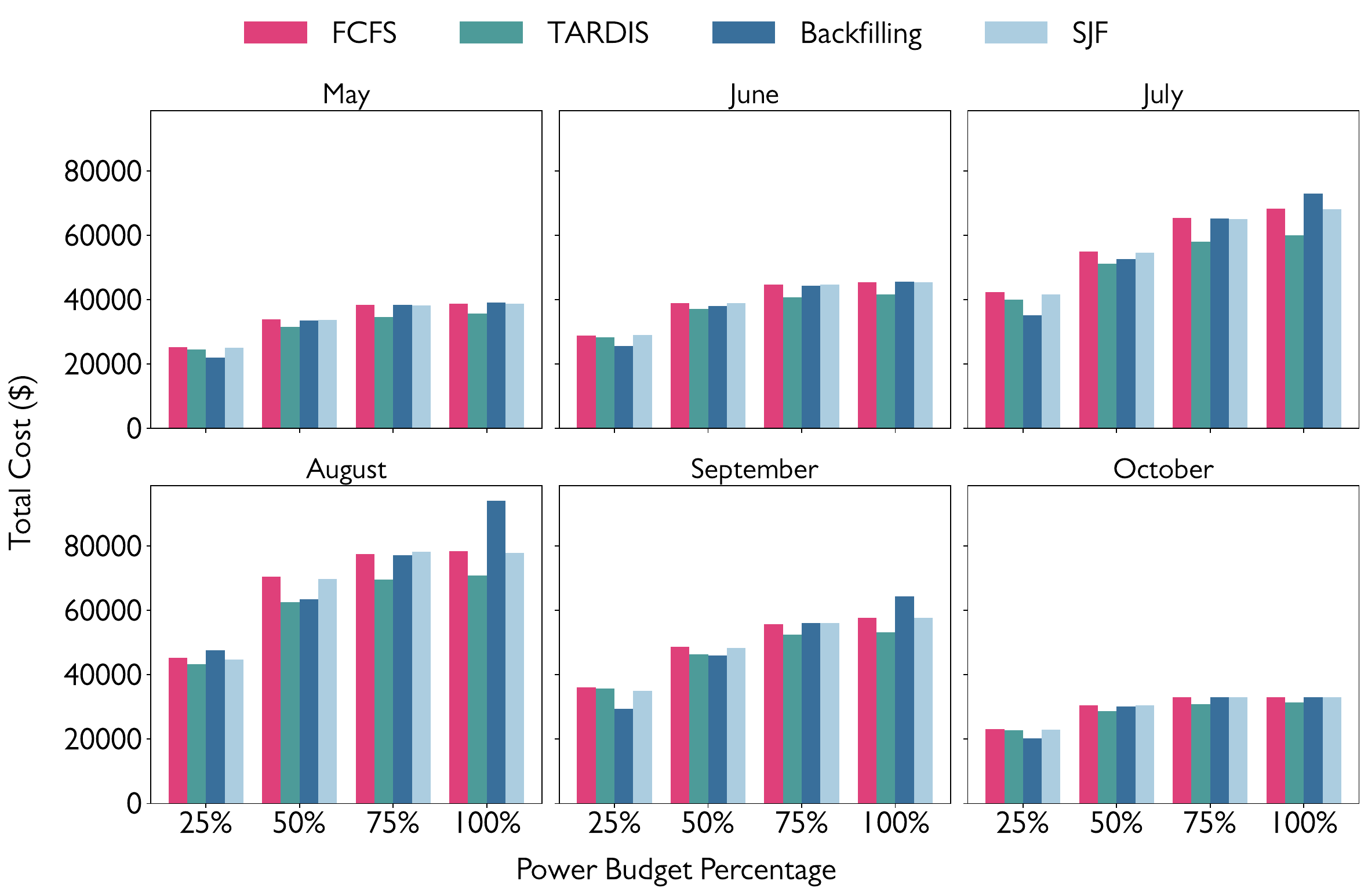}
    \caption{Total electricity cost.}
    \label{fig:total_cost}
\end{figure}

\begin{figure}[H]
    \centering
    \includegraphics[width=1\textwidth]{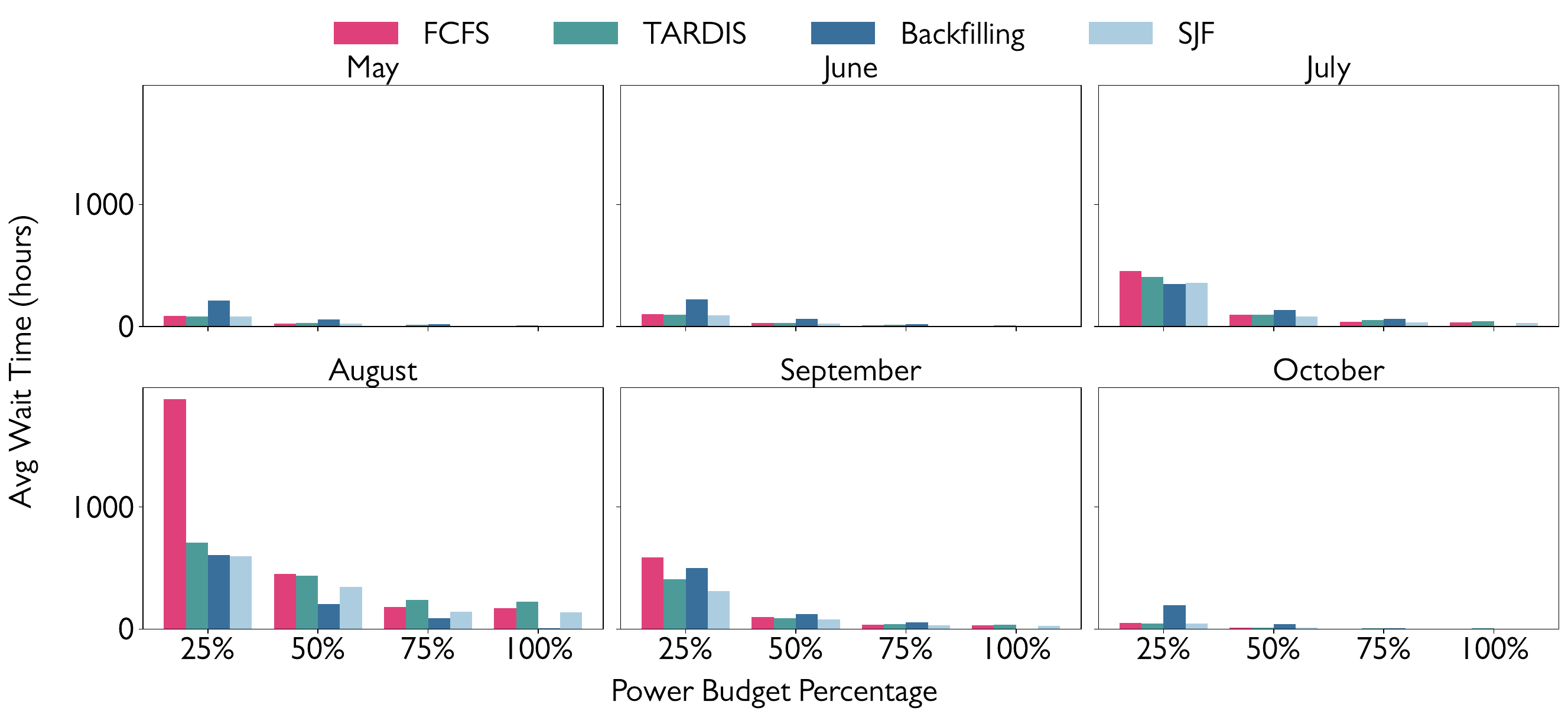}
    \caption{Average wait time.}
    \label{fig:wait_time}
\end{figure}

Our evaluation examines performance across varying power budgets (25\%, 50\%, 75\%, and 100\% of peak power consumption).
% to demonstrate how the scheduler leverages power capping during peak hours to reduce electricity costs. 
% By imposing power budgets, the scheduler can strategically defer high-power jobs to off-peak periods when electricity rates are lower, while still maintaining system performance and job throughput. 
Figure~\ref{fig:total_cost} shows the total electricity cost incurred by different schedulers across six months of operation. \ouralg consistently outperforms baseline schedulers across all power budget levels, with particularly strong performance during high-workload months (July-August). At 25\% power budget, the algorithm achieves its highest cost reductions by effectively shifting power-intensive workloads to off-peak hours.
Even at higher power budgets, \ouralg maintains its cost advantage by intelligently scheduling jobs based on their predicted power consumption patterns. 
% The cost improvement percentage relative to FCFS (Figure \ref{fig:cost_improvement_percent}) demonstrates that \ouralg effectively reduces electricity costs through strategic temporal job placement. 
During months with higher workload intensity (e.g., May-September), the scheduler achieves up to 18\% cost reduction by identifying and exploiting opportunities to run high-power jobs during off-peak hours. Figure~\ref{fig:wait_time} presents the average job wait times across different configurations. The results show that \ouralg's power-aware scheduling decisions do not significantly impact job wait times compared to traditional schedulers. While there is a slight increase in wait times under lower power budgets, particularly during high-workload periods, this represents a controlled trade-off between power cost optimization and job turnaround time. The scheduler maintains this balance by selectively deferring only those jobs whose power consumption patterns justify the delay in terms of overall cost savings.

\textbf{Peak vs. off-peak job distribution analysis:}

\begin{figure}[H]
    \centering
    \begin{subfigure}[b]{0.48\textwidth}
        \centering
        \includegraphics[width=\textwidth]{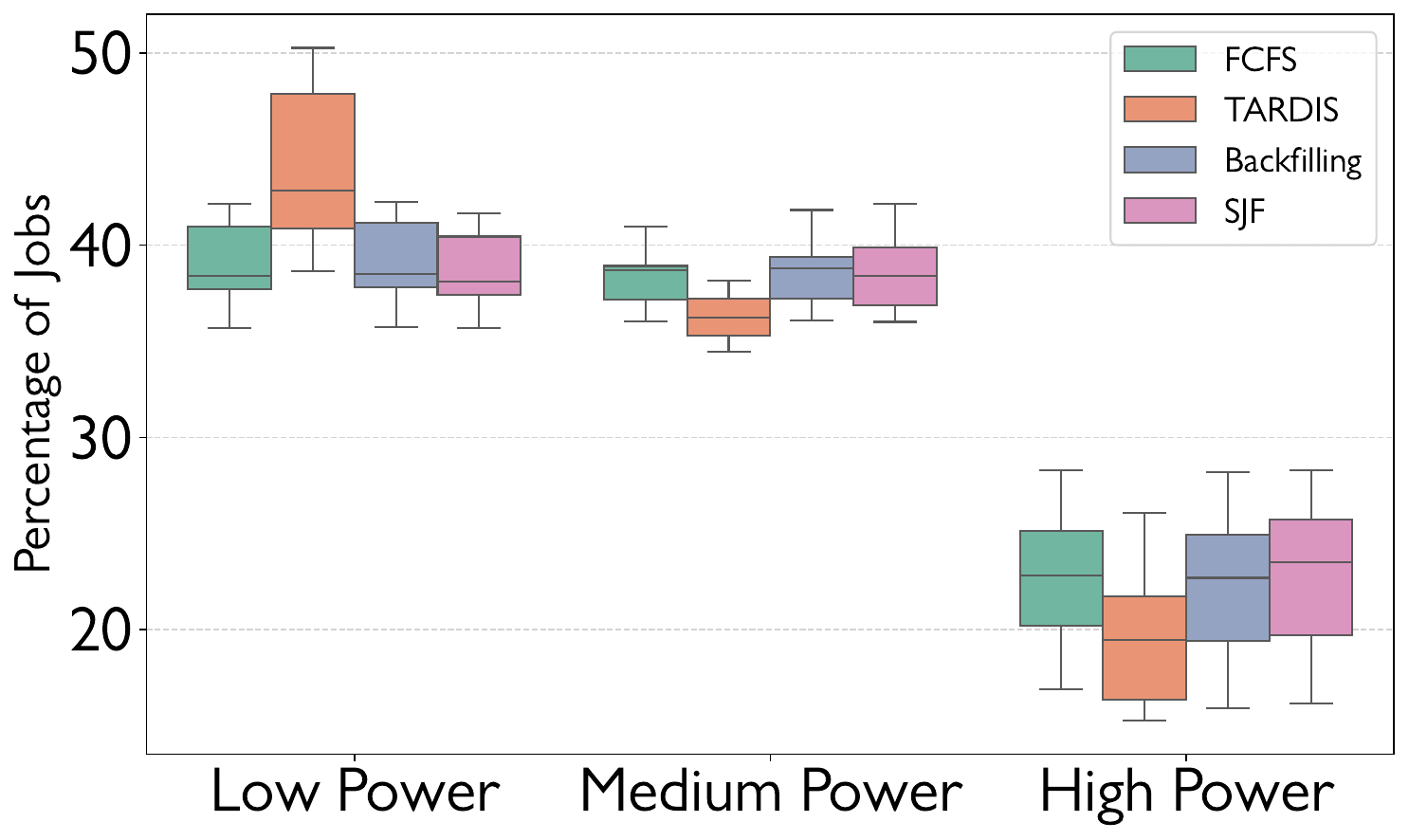}
        \caption{Peak hours.}
        \label{fig:peak_hours_distribution}
    \end{subfigure}
    \hfill
    \begin{subfigure}[b]{0.48\textwidth}
        \centering
        \includegraphics[width=\textwidth]{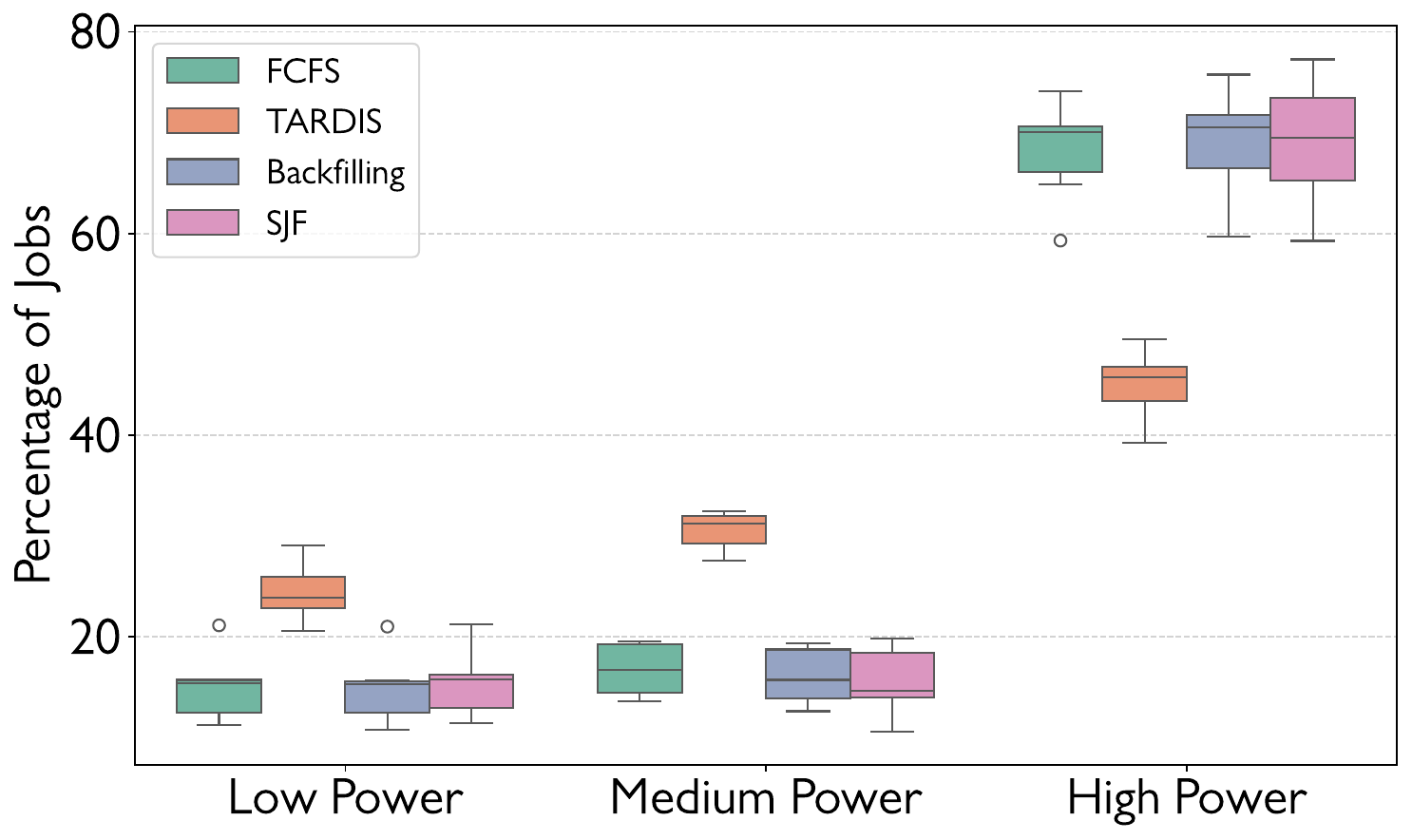}
        \caption{Off-peak hours.}
        \label{fig:off_peak_hours_distribution}
    \end{subfigure}
    \caption{Distribution of job execute by each scheduler during peak hours and off-peak hours.}
    \label{fig:peak_off_peak_distribution}
\end{figure}

We categorize jobs into low, medium, and high power groups based on their relative distribution in the dataset rather than absolute wattage thresholds. During peak hours (Figure \ref{fig:peak_hours_distribution}), \ouralg demonstrates a distinct scheduling pattern compared to baseline schedulers. While maintaining similar execution rates for low and medium power jobs, it significantly reduces the execution of high-power jobs to approximately 18\% compared to 22-25\% in other schedulers. This selective deferral of power-intensive jobs during expensive rate periods contributes directly to cost savings.  The complementary behavior is observed during off-peak hours (Figure~\ref{fig:off_peak_hours_distribution}), where \ouralg executes a notably higher percentage of high-power jobs (approximately 70\%) compared to baseline schedulers (45-50\%). This strategic shift of power-intensive workloads to periods of lower electricity rates showcases the scheduler's ability to make power-aware decisions. Meanwhile, the scheduler maintains balanced execution of low and medium power jobs during off-peak hours, ensuring efficient resource utilization while optimizing for electricity costs. 
% The results confirm that \ouralg's cost savings primarily come from its intelligent temporal placement of high-power jobs, rather than from any significant disruption to the execution pattern of lower-power workloads. The scheduler effectively leverages the power consumption predictions from the GNN model to make these strategic scheduling decisions.

\begin{figure}[H]
    \centering
    \includegraphics[width=0.9\textwidth]{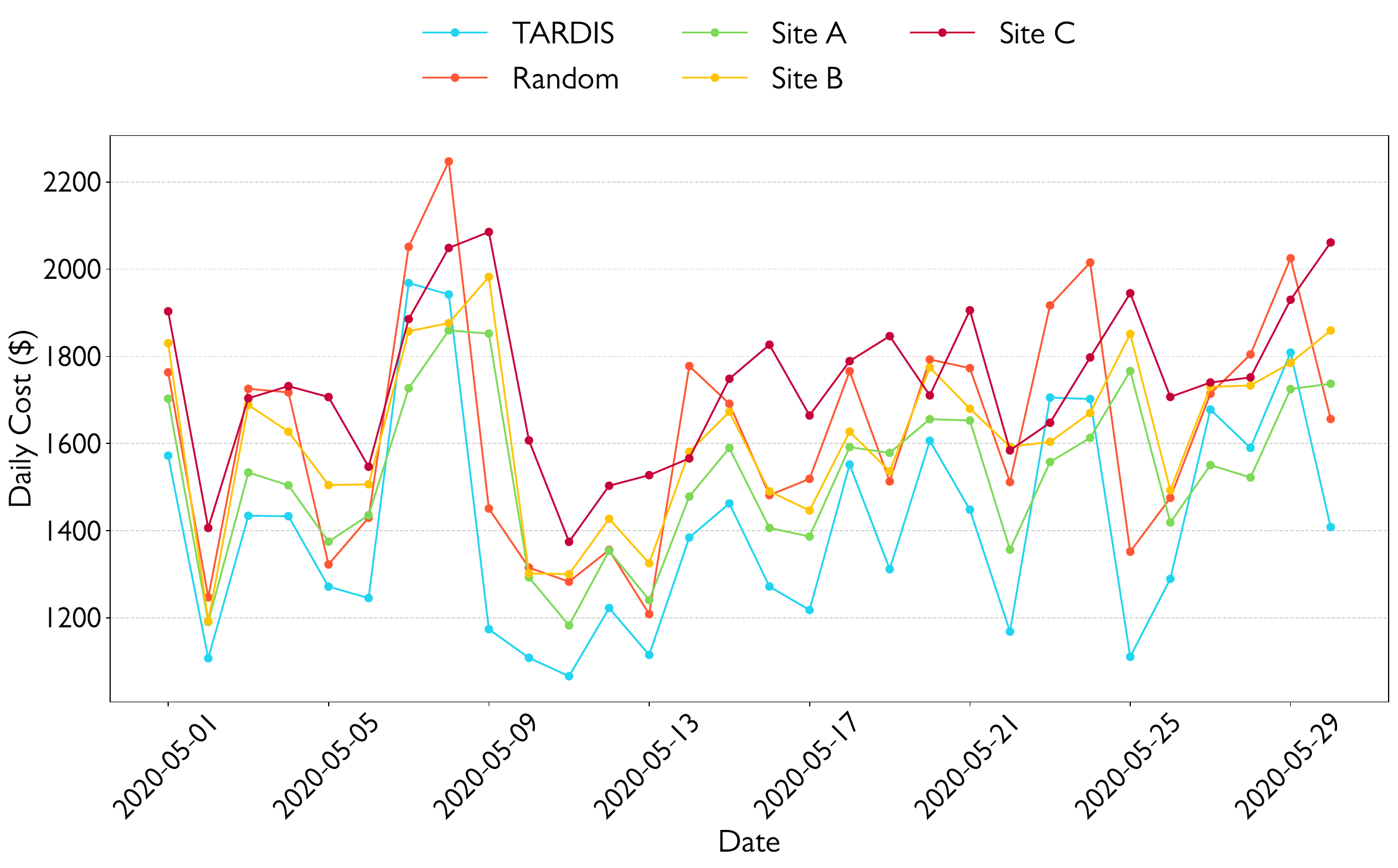}
    \caption{\ouralg achieves low average daily cost compared to single zone schedulers}
    \label{fig:multi_daily_cost_trends}
\end{figure}

\textbf{Multi-site optimization performance:}
The spatial dimension of our optimization approach is evaluated by comparing \ouralg against single-site schedulers and a random assignment policy across multiple HPC centers. Figure~\ref{fig:multi_daily_cost_trends} illustrates the daily cost incurred by different scheduling approaches over a month-long period. \ouralg consistently achieves lower daily electricity costs compared to other benchmarks. The daily cost for \ouralg ranges from approximately \$100 to \$650, significantly below the \$200-\$2,200 range observed for single-site schedulers (Site A, Site B, and Site C). This substantial difference demonstrates the effectiveness of our spatial-temporal optimization strategy, which leverages geographical electricity price variations to minimize overall costs.

The random assignment scheduler, which distributes jobs randomly across sites without considering power profiles or electricity rates, consistently performs worst with daily costs frequently exceeding \$1,800. Particularly notable are the cost spikes visible around May 9 and May 27, where all single-site schedulers experience significant cost increases, while \ouralg maintains relatively stable and low costs. This performance gap illustrates how \ouralg exploits the non-overlapping peak periods across different time zones.  Single-site schedulers (Site A, Site B, and Site C) show varied performance depending on their local electricity pricing, but all remain substantially more expensive than \ouralg. By intelligently routing power-intensive jobs to whichever site is currently experiencing off-peak rates, the scheduler achieves global optimization that would not be possible with localized scheduling decisions. 

\begin{figure}[H]
    \centering
    \includegraphics[width=1\textwidth]{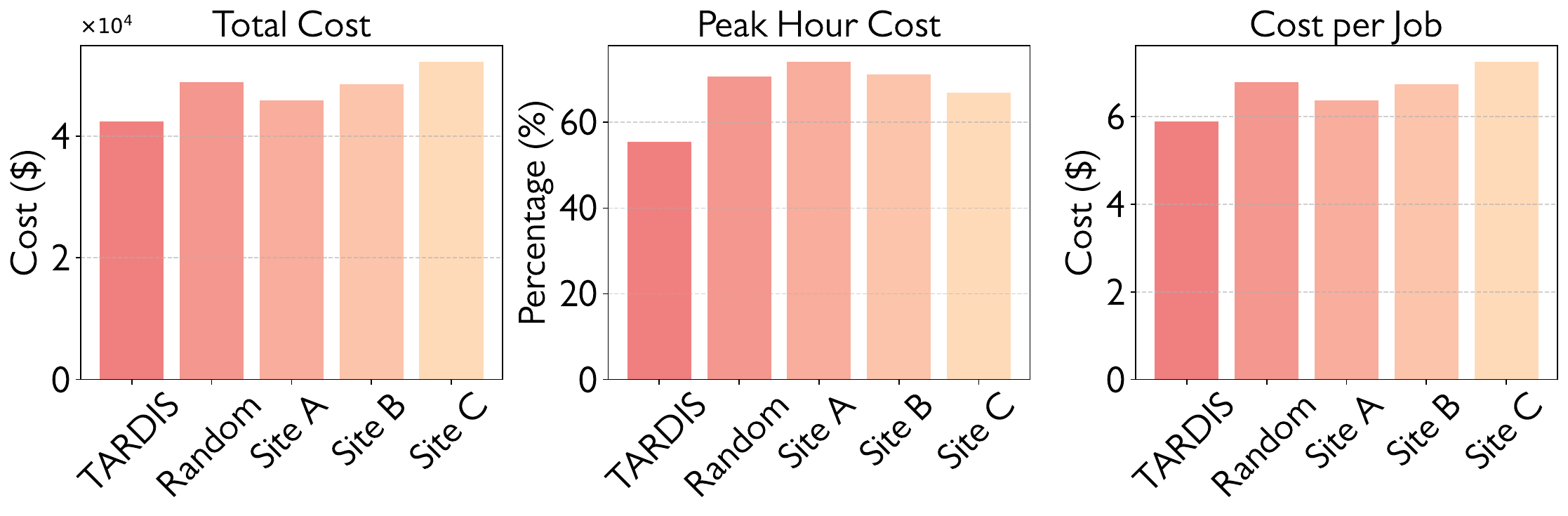}
    \caption{Aggregate cost comparison.}
    \label{fig:multi_total_cost}
\end{figure}

\textbf{Aggregate cost analysis:}
To quantify the overall impact of our multi-site optimization approach, we present an aggregate cost analysis in Figure~\ref{fig:multi_total_cost}. We compare \ouralg against single-site schedulers and random assignment across three key metrics (total cost, peak hour cost, cost per job). The total cost comparison (left panel) shows that \ouralg achieves the lowest overall electricity cost at approximately \$40,000 for the evaluation period, representing a 10-15\% reduction compared to the single-site schedulers (Site A, Site B, and Site C) and a 20\% reduction compared to random assignment. This significant cost saving demonstrates the cumulative benefit of making power-aware scheduling decisions across geographical locations. The peak hour cost analysis (middle panel) reveals one of the primary mechanisms behind \ouralg's efficiency. While single-site schedulers and random assignment execute 65-70\% of their workloads during peak hours, \ouralg reduces this percentage to approximately 55\%. This strategic shifting of workloads away from peak hours across multiple time zones allows the scheduler to minimize exposure to high electricity rates, without requiring excessive job delays. The cost per job metric (right panel) further confirms the efficiency of our approach, with \ouralg achieving the lowest per-job cost at around \$5.5, compared to \$6.5-\$7.0 for other approaches. This normalized metric indicates that the cost savings are consistent across jobs of different sizes and characteristics, rather than being driven by particular workload patterns or job types. Collectively, these results demonstrate that \ouralg effectively leverages both temporal and spatial dimensions to optimize electricity costs.

% By considering the predicted power consumption of jobs alongside the time-varying electricity rates across multiple geographic locations, the scheduler achieves substantial and consistent cost reductions without compromising system performance.

\begin{figure}[H]
    \centering
    \begin{subfigure}[b]{0.48\textwidth}
        \centering
        \includegraphics[width=\textwidth]{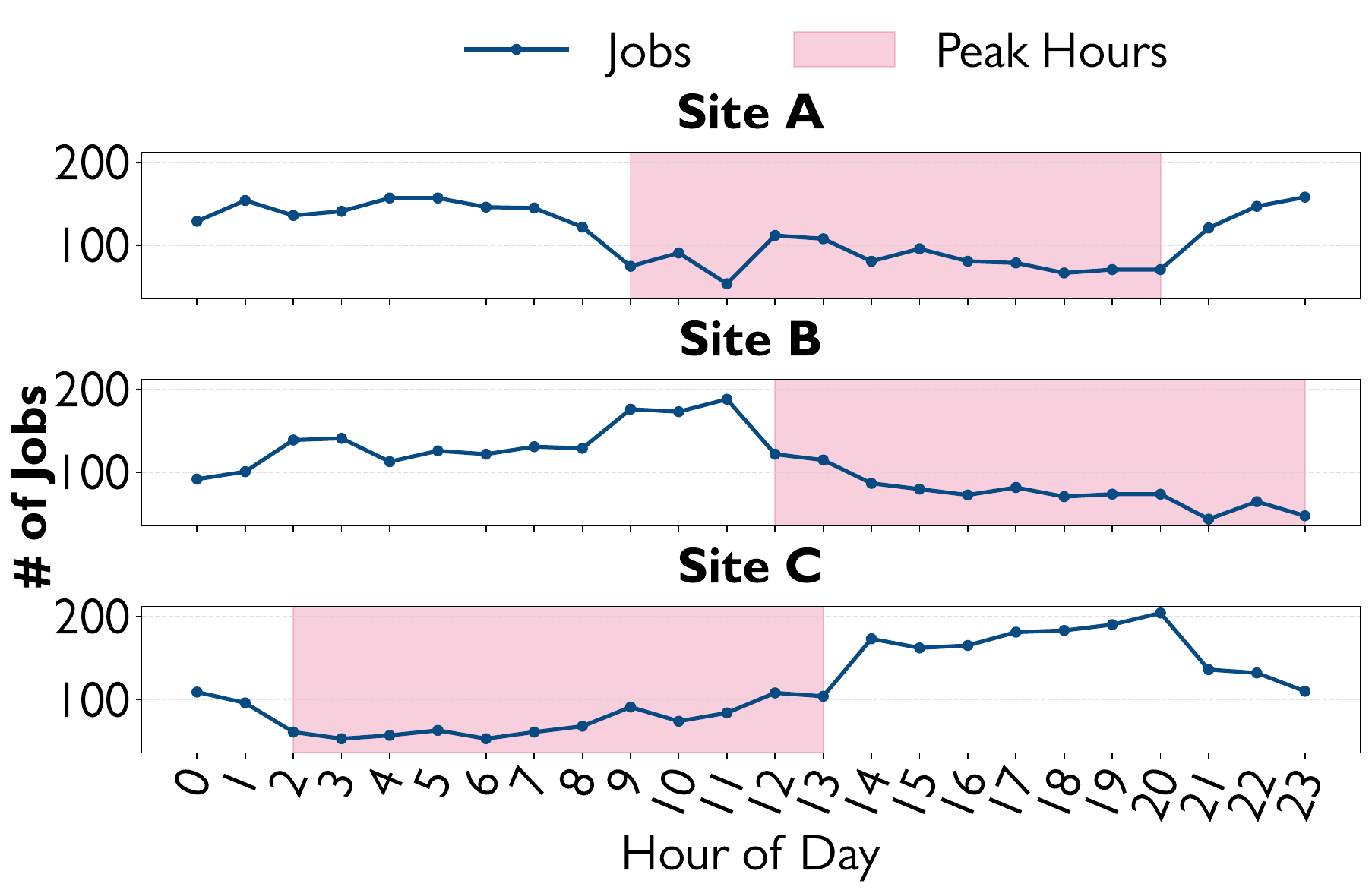}
        \caption{Job distribution.}
        \label{fig:multi_job_dist_peak_offPeak}
    \end{subfigure}
    \hfill
    \begin{subfigure}[b]{0.48\textwidth}
        \centering
        \includegraphics[width=\textwidth]{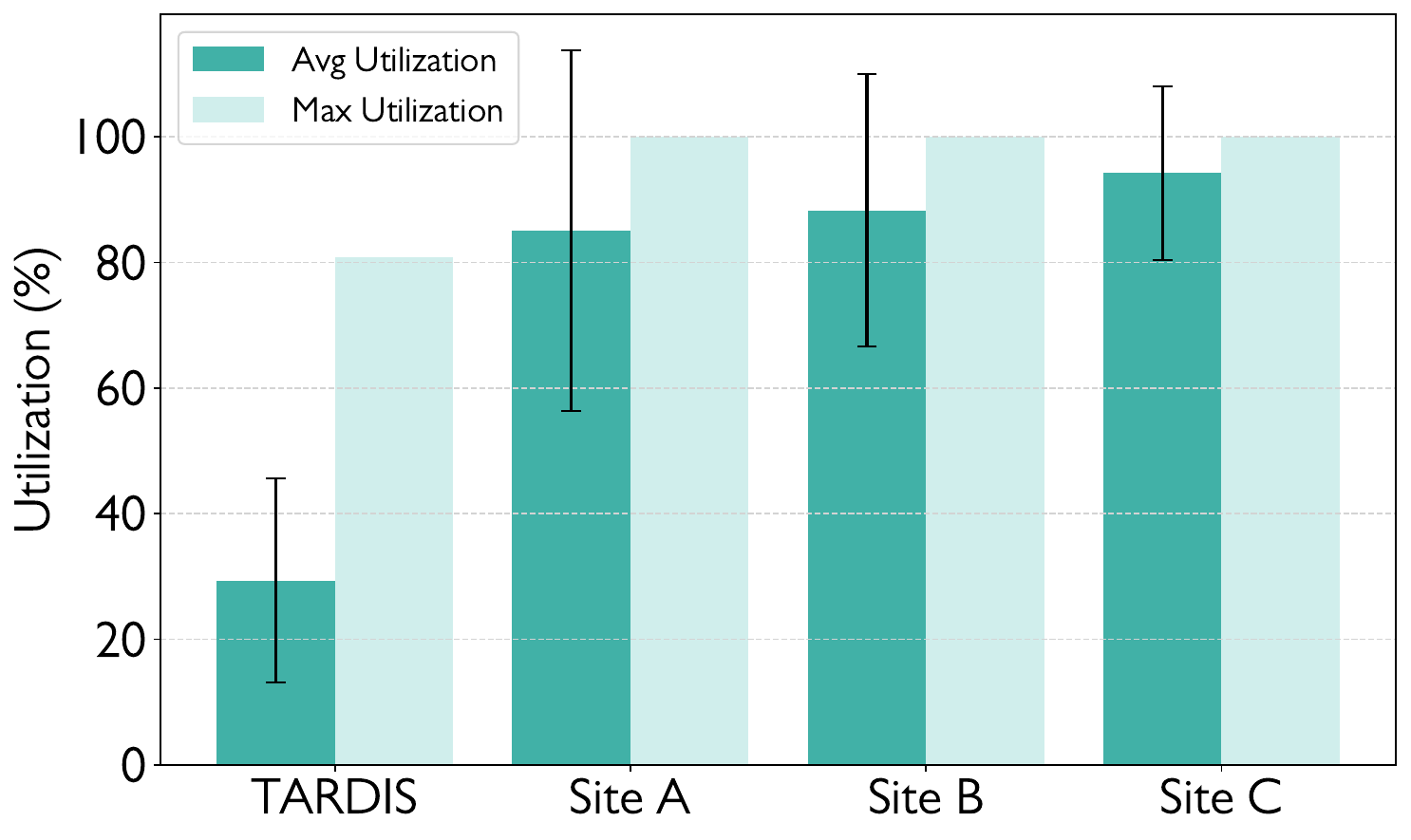}
        \caption{System utilization.}
        \label{fig:multi_avg_utilization}
    \end{subfigure}
    \caption{{Intelligent jobs distribution across multiple sites to achieve optimized system utilization.}}
    \label{fig:multi_job_dist_avg_utilization}
\end{figure}

\textbf{Job distribution and system utilization analysis.}
Figure~\ref{fig:multi_job_dist_avg_utilization} shows how \ouralg achieves cost optimization through strategic job distribution across multiple sites. The job distribution across hours of the day (Figure~\ref{fig:multi_job_dist_peak_offPeak}) reveals clear patterns in \ouralg's spatial-temporal scheduling decisions. For Site A, we observe relatively high job execution rates during non-peak hours (hours 0-6 and 22-23), with a noticeable decrease during peak hours (hours 6-22). This pattern indicates that \ouralg deliberately reduces workload at this site during its high-rate period. Site B shows a similar pattern but with a time shift due to its different time zone, with job counts dropping as the site enters its peak period. Particularly interesting is Site C, where job allocation significantly increases during hours 10-20, which correspond to peak hours for Sites A and B but off-peak hours for Site C. This demonstrates how \ouralg exploits the staggered peak periods across different time zones to minimize overall electricity costs. The system utilization metrics (Figure\ref{fig:multi_avg_utilization}) reveal another important aspect of \ouralg's efficiency. While single-site schedulers maintain high average utilization rates (80-95\%), \ouralg operates with a lower average utilization of approximately 30\%, but still achieves similar maximum utilization (80\% compared to 100\% for single sites). This deliberate underutilization provides the scheduler with flexibility to shift jobs between sites as electricity rates fluctuate throughout the day. Rather than maximizing utilization at each individual site, \ouralg optimizes utilization across the entire system in a way that minimizes total electricity costs. This analysis confirms that \ouralg's cost savings come from its ability to make intelligent job placement decisions that account for both temporal variations in electricity rates and the spatial distribution of computational resources. By maintaining flexibility in resource allocation and exploiting non-overlapping peak periods, the scheduler achieves significant cost reductions without compromising overall throughput.

\section{Conclusion}
\label{sec:conclude}
In this paper, we introduced a novel power-aware job scheduling framework for HPC systems that optimizes electricity costs by leveraging both temporal and spatial dimensions. Our approach combines graph neural networks for accurate power consumption prediction with a multi-objective scheduling algorithm that intelligently distributes jobs across time and space to minimize electricity costs while maintaining system performance. 
% The key contributions of our work include: (1) a GNN-based power prediction model that accurately estimates job-specific power consumption, (2) a temporal optimization algorithm that shifts power-intensive workloads to off-peak hours, and (3) a spatial optimization component that exploits geographical differences in electricity pricing across multiple HPC sites.
Our comprehensive evaluation using the PM100 dataset demonstrates that our scheduler consistently outperforms traditional approaches, achieving up to 18\% cost reduction in temporal optimization scenarios and 10-20\% savings in multi-site environments. The detailed analysis of job distribution patterns confirms that these savings come from intelligent workload shifting based on predicted power consumption and time-varying electricity rates. While our approach introduces a moderate increase in average job wait times, the trade-off is well-balanced, with the scheduler maintaining competitive system throughput. Furthermore, our analysis of system utilization reveals that the scheduler strategically maintains flexibility across sites to exploit rate differentials, rather than maximizing utilization at individual locations. 
% Future work could extend this approach to incorporate renewable energy availability predictions, further enhancing the environmental sustainability of HPC operations. Additionally, exploring federation mechanisms that allow jobs to migrate between sites could provide even greater optimization opportunities. As electricity costs and environmental concerns continue to grow in importance for HPC operations, power-aware scheduling approaches like \ouralg will become increasingly valuable for sustainable scientific computing.
%
%





% LNCS Homepage, \url{http://www.springer.com/lncs}, last accessed 2023/10/25
% \end{thebibliography}

\bibliographystyle{ieeetr}
\bibliography{ref}

\begin{thebibliography}{10}

\bibitem{inl2024mobile}
{Idaho National Laboratory (INL)}, ``Mobile supercomputer of the future: Inl researchers explore connecting data centers to microgrids \& microreactors,'' 2024.

\bibitem{enterpriseviewpoint2024sustainable}
{Enterprise Viewpoint}, ``Sustainable high-performance computing,'' 2024.

\bibitem{top500}
``{TOP500 List of the Fastest Supercomputers}.'' \url{(https://top500.org/)}.

\bibitem{miyazaki2018bayesian}
T.~Miyazaki, I.~Sato, and N.~Shimizu, ``Bayesian optimization of hpc systems for energy efficiency,'' in {\em ISC'18}, pp.~44--62, Springer, 2018.

\bibitem{nana2023energy}
R.~Nana, C.~Tadonki, P.~Dokl{\'a}dal, and Y.~Mesri, ``Energy concerns with hpc systems and applications,'' {\em arXiv preprint arXiv:2309.08615}, 2023.

\bibitem{borenstein2002dynamic}
S.~Borenstein, M.~Jaske, and A.~Rosenfeld, ``Dynamic pricing, advanced metering, and demand response in electricity markets,'' 2002.

\bibitem{santos2022understanding}
D.~C.~d. Santos, ``Understanding the energy consumption of hpc scale artificial intelligence,'' {\em arXiv preprint arXiv:2212.00582}, 2022.

\bibitem{yang2013integrating}
X.~Yang, Z.~Zhou, S.~Wallace, Z.~Lan, W.~Tang, S.~Coghlan, and M.~E. Papka, ``Integrating dynamic pricing of electricity into energy aware scheduling for hpc systems,'' in {\em SC}, pp.~1--11, 2013.

\bibitem{liu2013data}
Z.~Liu, A.~Wierman, and Y.~e.~a. Chen, ``Data center demand response: Avoiding the coincident peak via workload shifting and local generation,'' in {\em Proc. SIGMETRICS}, pp.~341--342, 2013.

\bibitem{haghshenas2020infrastructure}
K.~Haghshenas, S.~Taheri, M.~Goudarzi, and S.~Mohammadi, ``Infrastructure aware heterogeneous-workloads scheduling for data center energy cost minimization,'' {\em IEEE Transactions on Cloud Computing}, vol.~10, no.~2, pp.~972--983, 2020.

\bibitem{purkayastha2018holistic}
A.~Purkayastha, S.~Hammond, R.~Nagappan, and M.~Alt, ``Holistic approaches to hpc power and workflow management,'' in {\em IGSC}, pp.~1--8, IEEE, 2018.

\bibitem{fan2021job}
Y.~Fan, ``Job scheduling in high performance computing,'' {\em arXiv preprint arXiv:2109.09269}, 2021.

\bibitem{antici2023pm100}
F.~Antici, M.~Seyedkazemi~Ardebili, and A.~e.~a. Bartolini, ``Pm100: A job power consumption dataset of a large-scale production hpc system,'' in {\em SC'23 W}, pp.~1812--1819, 2023.

\bibitem{borghesi2017power}
A.~Borghesi, ``Power-aware job dispatching in high performance computing systems,'' 2017.

\bibitem{halder2024empirical}
D.~Halder, M.~Acharya, and A.~e.~a. Malsane, ``Empirical evaluation of ml for job power prediction,'' in {\em Proc. ACM/SPEC ICPE Companion}, pp.~181--188, 2024.

\bibitem{saillant2020predicting}
T.~Saillant, J.-C. Weill, and M.~Mougeot, ``Predicting job power consumption based on rjms submission data in hpc systems,'' in {\em ISC}, pp.~63--82, 2020.

\bibitem{zhou2014reducing}
Z.~Zhou, Z.~Lan, and W.~e.~a. Tang, ``Reducing energy costs for ibm blue gene/p via power-aware job scheduling,'' in {\em JSSPP'14}, pp.~96--115, 2014.

\bibitem{patki2015practical}
T.~Patki, D.~K. Lowenthal, A.~Sasidharan, M.~Maiterth, B.~L. Rountree, M.~Schulz, and B.~R. De~Supinski, ``Practical resource management in power-constrained, high performance computing,'' in {\em Proceedings of the 24th international symposium on high-performance parallel and distributed computing}, pp.~121--132, 2015.

\bibitem{mammela2012energy}
O.~M{\"a}mmel{\"a}, M.~Majanen, R.~Basmadjian, H.~De~Meer, A.~Giesler, and W.~Homberg, ``Energy-aware job scheduler for high-performance computing,'' {\em Computer Science-Research and Development}, vol.~27, pp.~265--275, 2012.

\bibitem{solorzano2024toward}
A.~L.~V. Sol{\'o}rzano, K.~Sato, and K.~e.~a. Yamamoto, ``Toward sustainable hpc: In-production deployment of incentive-based power efficiency mechanism on the fugaku supercomputer,'' in {\em SC24}, pp.~1--16, 2024.

\bibitem{pinheiro2001load}
E.~Pinheiro, R.~Bianchini, E.~V. Carrera, and T.~Heath, ``Load balancing and unbalancing for power and performance in cluster-based systems,'' tech. rep., Rutgers University, 2001.

\bibitem{sun2024energy}
J.~Sun, Z.~Gao, and D.~e.~a. Grant, ``Energy dataset of frontier supercomputer for waste heat recovery,'' {\em Sci. Data}, vol.~11, no.~1, p.~1077, 2024.

\bibitem{barroso2007case}
L.~A. Barroso and U.~H{\"o}lzle, ``The case for energy-proportional computing,'' {\em Computer}, vol.~40, no.~12, pp.~33--37, 2007.

\bibitem{fan2007power}
X.~Fan, W.-D. Weber, and L.~A. Barroso, ``Power provisioning for a warehouse-sized computer,'' {\em SIGARCH}, vol.~35, no.~2, pp.~13--23, 2007.

\bibitem{eia2024electricity}
{U.S. Energy Information Administration (EIA)}, ``Average retail price of electricity to ultimate customers,'' 2024.

\end{thebibliography}
\end{document}